\documentclass[preprint, amsmath,amssymb, aps, pre]{revtex4-1}

\usepackage{graphicx}
\usepackage{dcolumn}
\usepackage{bm}
\usepackage{mathrsfs}
\usepackage{hyperref}

\usepackage{float}
\usepackage{chemformula}

\begin{document}

\title{Role of symmetry in the orientationally disordered crystals of hard convex polyhedra}

\author{Sumitava Kundu}
\author{Kaustav Chakraborty}%
\author{Avisek Das}
\email{mcsad@iacs.res.in}%
\affiliation{School of Chemical Sciences, Indian Association for the Cultivation of Science, Kolkata, INDIA }%

\date{\today}

\begin{abstract}
The crystalline solids with lack of orientational ordering of anisotropic particles serve the purpose of studying the disordered systems with many fundamental applications in contemporary research. Despite the orientational disorder, multiple unique orientations with fixed angular differences exist in the crystal structures giving rise of ``discrete plastic crystal'' phase where the particles jump discretely within the unique orientations. We report the computational evidence of the role of symmetries between polyhedral particles and respective crystalline structures in controlling the existence of such phase at comparatively higher range of packing fractions beyond the freely rotating plastic crystals. The point groups of the particle and crystal structure were found to be directly connected in terms of the parallel alignment between the highest order rotational symmetry axes of the particle point group and any rotational axes of crystallographic point group, as a characteristic feature of this phase giving rise of discrete orientations. Based on our previous research [Kundu \textit{et al.}, arXiv:2311.06799, 2023] and new findings reported here, this symmetry relationship appeared to occur at the unit cells of the crystal structures which acted as the source of correlation, where as, all previously reported conserved orientational attributes i.e., number of unique orientations with fixed angular differences, equal population densities within the unique orientations, could be thought as the signatures of correlation present in the entire system. This relationship appeared to control all the aspects of phase which might be useful to draw fundamental insights about the disordered phases with orientational correlation as well as designing the disorder in the crystals.

\end{abstract}

\pacs{Valid PACS appear here}
\maketitle


\section{\label{sec:Introduction}Introduction} 
Ordering of classical many-body systems is a spontaneous process in nature which occur in different length scales \cite{Whitesides2002}. Crystallization drives a disordered system to a translationally ordered state where all the constituents form a translational broken symmetry state as observed in the domain of colloidal crystals \cite{Pawel1983, Dinsmore1998, Li2016, Orr2022, Boles2016}, molecular crystals \cite{Timmermans1961, Mccullough1961, Reynolds1975, Harfenist1996, Loidl1990, Nitta1959} or nanoparticle superlattices \cite{Mirkin1996, Robert1996, Whitesides2002, Maye2007,Henzie2012, Bodnarchuk2011, Boneschanscher2014,  Haixin2017, Zhou2022, Geuchies2016, Shevchenko2006, Zhang2013, Jones2010, Glotzer2007}. Crystal structures of anisotropic particles or molecules exhibit different kinds of orientational phases due to the rotational motion of the constituents while maintaining proper translational order \cite{Timmermans1961, Nitta1959, Reynolds1975, Mirkin1996, Glotzer2007, OlveraDeLaCruz2016, Meijer2017, Damasceno2012c, Gantapara2013}. Understanding the orientations of the molecules or anisotropic particles would serve the purpose to achieve various kind of targeted materials with desired properties or functionalities \cite{Nitta1959, Whitesides2002, Glotzer2007}.

The rotational motion of the building blocks led to the possibility of different orientational phases in the crystalline assemblies which include the plastic crystals \cite{Timmermans1961, Reynolds1975, Lee2023, Karas2019, Dullens2007, Meijer2017}, orientationally ordered crystals \cite{Damasceno2012c, Karas2019, Gantapara2013, Lu2021, Henzie2012}, correlated disordered crystals \cite{Keen2015,Chaney2021,Meekel2021} or orientational glass \cite{Loidl1989, Loidl1990, Karas2019, Abbas2022, Vdovichenko2015}. Rotator or plastic crystal is a orientationally disordered phase with a long-range translational order and short-range orientational order \cite{Timmermans1961, Reynolds1975} occurring at very lower range of packing fractions. The existence of ``plastic crystal'' is well documented in the field of colloidal crystals \cite{Pawel1983, Orr2022, Boles2016}, molecular crystals \cite{Vdovichenko2015, Even2016, Beake2017, Loidl1990} or nanocrystal superlattices \cite{Shevchenko2006, Zhang2013, Damasceno2012b, Maye2007, Lee2023, Damasceno2012c}. A number of studies referred to such states as ``freely rotating plastic crystal'' \cite{Edington1999, Liu2014, Meijer2017, Harada2021, Karas2019, Damasceno2012c, Lee2023}, where as, few anisotropic orientational choices were also observed in the plastic crystal under the influence of hard-core interaction \cite{Torquato2009, Gantapara2013, Gantapara2015a, Sharma2024}. The existence of orientationally ordered crystal structure with the long-range positional and orientational order \cite{Damasceno2012c, Lu2021, Henzie2012} indicated another extreme behavior of the anisotropic particles compared to the completely random orientational phase in the crystal structures. Direct experimental observations of orientational order and disorder were reported in colloidal crystals \cite{Dullens2007, Meijer2017} and nanocrystal superlattices \cite{Deng2020,Elbert2021,Abbas2022} very recently. A detail realization of the nature of disorder turned out to be an intriguing field to appreciate the orientational arrangements of the particles in the crystalline solids. The versatile orientational behavior of the particles maintaining the translational order became one of promising fields in the contemporary research to fabricate real-life applications.  

Entropy driven ordering became one of the promising fields in studying crystallization where the constituents interact via hard-core interaction only \cite{Onsager1949, Alder1957, Berryman1983,Frenkel1984,Frenkel1985,Veerman1992a, McGrother1996, Bolhuis1997}. To realize the direct experimental observations of various orientational phases in the crystalline solids, hard anisotropic particles were used as a promising model of colloidal and nanoparticles. A lot of computational studies have shown the evidences of the assembly of hard particles into different kinds of simple and complex structures \cite{Alder1957, Bolhuis1997, John2008, Haji-Akbari2009, Torquato2009, Damasceno2012c, Lee2023} reminiscent to the assembled states observed in the experiments of colloids and nanoscience \cite{Alloyeau2009, Abbas2022, Gong2017, Even2016, Ong2017, Meijer2017}. The accomplishment of achieving a variety of complex behavior in the assembled states of hard polyhedral particles compelled the study of this idealized model to investigate further. Different kinds of orientational behavior were investigated thoroughly using the hard polyhedral systems giving rise of all possible orientational phases \cite{Gantapara2015a, Shen2019, Karas2019, Damasceno2012c, Lee2023, Karas2019} as observed in the colloidal crystals, molecular solids or nanocrystal superlattices. Studying the orientational disorder in the crystals of hard polyhedra have been brought under scrutiny to serve the purpose of understanding the disordered crystalline structures in a more deeper way \cite{Torquato2009, Agarwal2011a, Damasceno2012b, Damasceno2012c, Gantapara2013, Gantapara2015a, Karas2019, Lee2023, Sharma2024}. In a recent study, we discussed about the existence of orientationally disordered crystalline phase in multiple polyhedral systems obtained from entropy driven assembly where the polyhedra were found to maintain certain discrete orientations while staying at the lattice sites depicted by the face centered cubic (FCC) crystals \cite{Kundu2023}. The existence of discrete orientations of the particles were reported in the crystal solids in both experiments \cite{Abbas2022} and computer simulations \cite{Shen2019, Karas2019, Lee2023, Gantapara2015a, Sharma2024}. Various investigations also revealed the discrete orientational hopping of the particles in both two \cite{Shen2019} and three dimensions \cite{Lee2023} where multiple unique orientations existed in the systems. In those studies, this phase was termed as ``discrete rotator phase'' or ``discrete plastic crystal''. The investigators pointed out that the essential differences of this phase and freely rotating plastic crystals could be realized in the context of ``discrete'' and ``continuous'' rotational motion of the particles respectively while maintaining the translational order. Our computational study showed the identical behavior of the particles at high density solids and characterized this phase as equilibrium in nature \cite{Kundu2023}. Multiple orientational attributes along with the number of unique orientations remained preserved in the entire system irrespective of the frozen orientational state at very high packing fractions or discretely mobile states at comparatively lower packing fractions. Proper characterization of multiple unique orientations at the unit cells revealed the crystalline systems as orientationally disordered solids. We argued the existence of long-range orientational behavior in the system where the particles of each unique orientations were present almost homogeneously in the entire crystal. Based on our analyses, we anticipated this phase to preserve the signatures of both long-range correlation and orientational disorder which were manifested in terms of various conserved attributes in the entire system. We empirically hypothesized the ``symmetry mismatch'' between the particle and crystal structure where the particles were oriented in such way, it broke few symmetries of the corresponding crystal structure. However, the existing literature also indicated the symmetry relationship between the point groups of the particles and local environments \cite{Shen2019, Lee2023} where the existence of continuous and discrete rotator phase were interpreted in the context of commonality of the operations in the corresponding point groups. It was explained that the particles were orientated in the assembled structures such a way, certain symmetry elements of the crystals were recovered by the particles due to the mismatch of specific elements in the corresponding point groups. This gave rise of an obvious question whether any straightforward relationship between the symmetries of the particles and crystal structures could be realized in order to explain the discrete orientational phase in the disordered solids in a much better way and this demands further attention to understand the essence of this phase thoroughly.

In this article, we present the computational study of the single-component systems of hard convex polyhedra to investigate the role of symmetry of the particle and crystal structure in the presence of orientational correlation. The chosen polyhedra were Elongated Pentagonal Dipyramid (EPD), Elongated Square Gyrobicupola (ESG) and Elongated Pentagonal Gyrocupolarotunda (EPG) which were reported to have multiple unique orientations in the discrete rotator phase in the FCC structures \cite{Karas2019, Kundu2023}. The freely rotating plastic crystal phase were also present for all the shapes at lower packing fractions before the transition of the crystals into the isotropic liquid phase. Though the detail phase behavior of these polyhedra were studied in earlier investigations, we used these shapes to identify some additional aspects in terms of the symmetry relationship between the particles and crystals. In this research, we explore the interplay between the point groups of the particle and crystal structure in controlling the unique orientations in the discrete rotator phase and observe an explicit relationship between the symmetry elements of the corresponding point groups. This direct relationship occurs at the unit cells of the crystal structure and appear to be the source of long-range correlation which satisfies the existence of all orientational attributes as the signatures of this correlation observed in this phase. Our findings might play an important role in developing fundamental theory for complex orientational broken symmetry states in the many-body system or understanding the orientational disorder from different aspects.

\section{Models and methods} \label{methods}
Three convex polyhedral shapes namely Elongated Pentagonal Dipyramid (EPD), Elongated Square Gyrobicupola (ESG) and Elongated Pentagonal Gyrocupolarotunda (EPG) were used in this study. The shapes are displayed as the insets in the first row of Figs.\,\ref{fig:snapshots_corr_diosorder}A,B,C. All these shapes were reported to form face-centered cubic (FCC) crystals in the entire solid region \cite{Damasceno2012c, Kundu2023}. The detail translational and orientational analyses were carried out in our previous investigation \cite{Kundu2023}. Multiple unique orientations appeared in the FCC crystals and the numbers were four, six and eight for EPD, ESG and EPG shapes respectively. Here, we used the same shapes to explore the unique orientations from the symmetry of both the particle and crystal structure. Three shapes were different from each other both from the geometry as well as point group symmetry. The EPD and ESG shapes have 12 vertices, 15 faces and 24 vertices, 26 faces respectively where the point groups are $D_{5h}$ and $D_{4d}$ point group respectively. The EPG shape is consists of 35 vertices and 37 faces and contains $D_{5h}$ point group symmetry. Single component systems of these three polyhedra were simulated using HOOMD-Blue simulation toolkit \cite{Anderson2016a}. We identified the unit cell of the crystal structures confirming the observations of the previous results \cite{Damasceno2012c, Kundu2023}. The point groups of the particles and corresponding crystal structures were analyzed thoroughly to realize the occurrence of distinct orientations of the particles located at the lattice sites of the unit cells depicted by the crystal structures.

\subsection{Simulation protocol} \label{sim_protocol}
We performed hard particle Monte Carlo simulations (HPMC) using HOOMD-Blue toolkit \cite{Anderson2016a}, where the metropolis method was determined by the overlapping of the particles . A dilute system was prepared followed by slow volume compression of the system up to $\phi$ $\sim$ 0.55 or 0.6, depending on the shapes. After the equilibration performed in constant volume simulation for sufficiently large Monte Carlo steps, the crystal structure was obtained. The reduced pressure $p^{\ast} = \beta p v_0$, where $\beta = (k_{B} T)^{-1}$ and $v_0$ is particles volume, was estimated from the constant volume crystallization simulation using the \textsl{scale distribution function} implemented in HOOMD. The particle volume $v_0$ was set to 1.0 to compare the system density with packing fraction. The system was compressed by increasing the pressure value slowly and allowed to equilibrate for 100 Monte Carlo (MC) steps under isobaric-isothermal ensemble (\textsl{NPT}). This protocol helped us to achieve the system at very high packing fraction after which no significant changes in system were observed. The entire compression simulation was performed at $\sim$ 25-30 pressure values consecutively where the system at highest achievable packing fraction was chosen for the melting simulations. We carried out the melting simulations under \textsl{NPT} ensemble at the same pressure values which were used during the compression runs. This was continued until the isotropic liquid phase was achieved. The simulation method allowed us to analyze the translational order of the structures and particle orientations at each frame of the packing fractions during the compression or expansion run of the system. Here, the data from the melting simulations were chosen to report the translational and orientational behavior of the particles. 

\subsection{Orientational difference and detection of the unique orientations} \label{unique_orein}
Single particle orientation is represented by quaternion in the system. The orientational difference between two particles is described as the minimum of all angles $\theta_{ij} = 2 \cos^{-1}[\Re(\mathcal{Q}^{\dagger}_{i} \mathcal{Q}_{j})]$, where $\mathcal{Q}_{i}$ and $\mathcal{Q}_{j}$ were the quaternions of particle $i$ and $j$ respectively and $\theta_{ij}$ was the quaternion angle. The proper rotational operations of the particle point group ($\Gamma^{p}$) was considered while calculating the pairwise angles between any two particles. Two particles are said to be perfectly ordered if $\theta_{ij}$ = 0. The distribution of all pairwise angles in the system was calculated measuring the overall angular dispositions of the particles but it failed to disclose the number of unique orientations in the system. To identify the unique orientations $\mathbb{Q}_k$, where $k = 1,2,\ldots,N_{\Omega}$ and $N_{\Omega}$ was the total number of unique orientations, a previously introduced recipe was followed by choosing three arbitrary quaternions form the system $\mathcal{Q}_{ref, 1}$, $\mathcal{Q}_{ref, 2}$, $\mathcal{Q}_{ref, 3}$ maintaining finite orientational differences \cite{Kundu2023}. For the $i$-th particle chosen from the system, three angles with the references corresponded to a set consisting of the angles $\Theta_{1, i}, \Theta_{2, i}, \Theta_{3, i}$. In this way, the sets of three angles were obtained for all particles in the system resulting a three dimensional distribution of the orientational differences. In presence of statistical noise, the distribution produced multiple clusters in the angular space where the total number of clusters indicated the number of unique orientations $N_{\Omega}$ in the system. The particles belonging to a same cluster were considered as orientationally ordered within the tolerance $\theta_{c}$ and the particles from different clusters were accounted as orientationally disordered.  

\subsection{Rotational axes of particle point group and crystallographic point group} \label{symm_relation}
In order to analyze multiple unique orientations of the particles sitting at the lattice sites of the crystal, calculation of the spatial arrangements of the orientations was required to know. To investigate the translational order, we detected the unit cell of the crystal structure. As the polyhedral particles self-assembled into a particular crystal structure, the point group symmetries of the particle and corresponding crystal were considered to explore the symmetry relationship between these two entities. Instead of the comparison between the rotational operations present in the point groups only, the axes of rotational symmetry operations corresponding to both the particle point group and the crystallographic point group were considered to decipher any relationships. In the three dimensional space, a vector indicates a unique direction and the alignment between two vectors can be realized by the intermediate angle. If the angle between the two directions is zero then the vectors appear to be exactly parallel; any finite value of the angle suggests the misalignment of the two vectors. This idea was used to know the alignments of differently oriented particles sitting at the lattice sites of the FCC unit cells. The directions of axes corresponding to the proper rotational symmetry operations of the particle and crystallographic point groups were measured followed by the comparison among these two sets of axes. To accomplish any direct correspondence, it was required to identify the orientation of the unit cell in the simulation box as it did not commensurate with the crystal structure obtained in the self-assembly upon compression of the isotropic phase. We evaluated the orientation of a unit cell which was defined by an orthogonal rotation matrix $\mathcal{R}_{c}$ in the global frame. Then, we determined the set of all rotational symmetry operations (both the rotational axes and angles) of the corresponding crystallographic point group $\Gamma^{c}$ in the local frame of unit cell. All the axes of the rotational symmetry operations corresponding to the particle point group $\Gamma_{p}$ were also obtained in the local frame of the particle. The rotational axes of the particle point group were further categorized according to the ``highest order'' and ``lower order''. The ``highest order'' rotational symmetry axes of the particle explicitly meant that if the particle point group contains few proper rotational operations of $C_2$, $C_3$ and $C_4$, then the axes corresponding to the $C_4$ operation are said to be ``highest order'' axes of the particle. All other axes except the ``highest order'' were referred to as ``lower order'' axes for the particle. The highest order rotational symmetry axes of the particle were defined as a set $\mathcal{A}^{p}_{max}$ where as, the lower order were considered as $\mathcal{A}^{p}_{min}$. Again, all the rotational symmetry axes of the corresponding crystallographic point group were defined as a set $\mathcal{A}^{c}$ in the corresponding local frames without any classification.

To formulate any correspondence among the rotational symmetry axes of the particle point group and crystallographic point group, the reference frame was required to be fixed for all the axes. Considering the global frame as reference, we carried out the orthogonal transformations from the local frame of the particle and corresponding unit cell to the global frame using the Equations \ref{eq:particle_max_global}, \ref{eq:particle_min_global}, \ref{eq:uc_global};
\begin{equation}\label{eq:particle_max_global}
	\mathcal{S}^{p}_{max, m, i} = \mathcal{R}_{i}\mathcal{A}^{p}_{max, m}
\end{equation}
\begin{equation}\label{eq:particle_min_global}
\mathcal{S}^{p}_{min, t, i} = \mathcal{R}_{i}\mathcal{A}^{p}_{min, t}
\end{equation}
\begin{equation}\label{eq:uc_global}
	\mathcal{S}^{c}_{n} = \mathcal{R}_{c} \mathcal{A}^{c}_{n}
\end{equation}
where $\mathcal{R}_{i}$ was the orthogonal rotational matrix defined in the global frame corresponding to the unit quaternion $\mathcal{Q}_{i}$ of $i$-th particle, as the quaternion is another representation of orthogonal matrix in three dimensions. $\mathcal{A}^{p}_{max, m}$, $\mathcal{A}^{p}_{min, t}$ and $\mathcal{A}^{c}_{n}$ were one of the elements of the sets $\mathcal{A}^{p}_{max}$, $\mathcal{A}^{p}_{min}$ $\mathcal{A}^{c}$ in the local frames respectively with $m$ $\in$ [$1$, $\mathcal{N}^{p}_{max}$], $t$ $\in$ [$1$, $\mathcal{N}^{p}_{min}$] and $n$ $\in$ [$1$, $\mathcal{N}^{c}$]. $\mathcal{N}^{p}_{max}$, $\mathcal{N}^{p}_{min}$ and $\mathcal{N}^{c}$ corresponded to the total number of highest order and lower order rotational axes of the particles and all rotational symmetry axes of the unit cell respectively. $\mathcal{S}^{p}_{max}$, $\mathcal{S}^{p}_{min}$ and $\mathcal{S}^{c}$ were the sets containing the highest order and lower order rotational axes of the particle and all rotational axes of the crystal respectively, defined in the global reference frame. The set $\mathcal{S}^{c}$ corresponding to all rotational axes of the crystal symmetry was classified according to the particular kind of proper rotational operation in the crystallographic point group i.e., $C_s$, where $s$ denoted the $s$-\textit{fold} rotation allowed in the crystallographic point group. For example, the ``cubic' crystal with $O_h$ symmetry contains only $C_2$, $C_3$, $C_4$ rotational operations where $s$ has three integer values; 2, 3 and 4. For each particle $i$, any rotational symmetry axis of the particle ($\mathcal{S}^{p}_{max}$ or $\mathcal{S}^{p}_{min}$) irrespective of the ``highest order'' or ``lower order'' was compared to the axes of set $\mathcal{S}^{c}$ based on certain kind of rotational operation $C_s$. The angles between the axes of two entities (particle and crystal) were calculated for all the particles in the system for each kind of rotational operation $C_s$. It should be noted that $\alpha_{Cs}$ denoted the angles between two axes. The distributions of angles $\alpha_{Cs}$ were plotted for all particles to observe the particular kind of crystallographic axes which appeared to be relevant in the alignments with the particle axes. Here, all the axes corresponding to a particular crystallographic rotational operation $C_s$ were considered as equivalent. These distributions were formulated to capture the alignments between any rotational axes of the particle and the  particular kind of rotational axes of the crystal based on the operation $C_s$. 

Next, we considered all the $\mathcal{S}^{c}$ crystal axes without any classification depending on the rotational operation $C_s$. All the axes were considered as distinguishable and indexed as integers followed by the identification of identities of $\mathcal{S}^{c}$ axes for particular operation $C_s$. We noted those identities of the crystallographic axes. For $i$-\textit{th} particle, the angle $\alpha_{m, i}$  was calculated as the minimum of all the angles formed between the $m$-th axis of the set $\mathcal{S}^{p}_{max}$ and all rotational axes of the crystal $\mathcal{S}^{c}$ irrespective of any kind of operation, according to the Equation \ref{eq:min_angle}.
\begin{equation}\label{eq:min_angle}
	\centering
	\alpha_{m, i} = \underset{n}{\text{\textit{min} }} \{\alpha_{m, n, i}\} = \underset{n}{\text{\textit{min} }} \{\cos^{-1}(\hat{\mathcal{S}}_{max, m, i}^{p} \cdot \hat{\mathcal{S}}_{n}^{c})\}
\end{equation}
where $m$ and $n$ were the indices of highest order rotational axes of the particle
and specific kind of rotational axes of the crystal responsible for the alignments which were chosen from the sets $\mathcal{S}^{p}_{max}$ and $\mathcal{S}^{c}$ respectively. For $i$-th particle, we could obtain a set of angles $\alpha$ consisting of all $\alpha_{m, i}$ values depending of the number of $\mathcal{S}^{p}_{max}$ axes. Each value of the set $\alpha$ was the angle between two vectors chosen from these two sets and could be thought as an ``axis-angle''. It should be noted that, the values of $\alpha$ would correspond to any particular kind of angles $\alpha_{Cs}$, if one specific kind of crystallographic axes appear to be responsible in such alignment. In some sense, the definition of $\alpha$ was specifically dependent on the highest order rotational axes of the particles and particular kind of responsible crystallographic axes and varied with systems. $\hat{\mathcal{S}}_{max, m, i}^{p}$ and $\hat{\mathcal{S}}_{n}^{c}$ denoted the unit vectors from the corresponding sets. The minimum angle $\alpha_{m, i}$ was calculated for all the particles in the system ($i = 1,2,\ldots,N$) which was defined as a set $\alpha_{m}$ and $N$ corresponded to the total number of particles in the system. The data of $\alpha_{m}$ was presented as a two-dimensional histogram to analyze the probabilities of particle alignments at the unit cell of the crystal structure. Simultaneously, the index $n$ of the set $\mathcal{S}^{c}$ was monitored for each particle $i$ to observe the alignment of the particle with any particular kind of rotational axis of the crystal based on the operation $C_s$. 

As the definition of $\alpha_{m}$ suggested, the effect of any angles from $\alpha_{C2}$, $\alpha_{C3}$, $\alpha_{C4}$ reflected in the distribution of $\alpha_{m}$, if any particular kind of rotational operation ($C_s$) of the crystallographic point group was responsible for the alignments. If the axes of $\mathcal{S}^{p}_{max}$ made the parallel alignment with the axes of $\mathcal{S}^{c}$, the distribution of $\alpha_m$ was expected to produce the peak near $\sim$ $0^{\circ}$. So, the existence of multiple unique orientations was perceived in the context of the alignments by comparing the directions of axes from the sets $\mathcal{S}^{p}_{max}$ and $\mathcal{S}^{p}_{min}$ with $\hat{\mathcal{S}}^{c}$. Considering the formulation, for a crystalline state with completely random orientations of the particles, the distribution profile was expected to produce a Gaussian peak at sufficiently far away from $0^{\circ}$ in the equilibrium phase. Considering the distribution as a reference state, any significant deviation from the reference might serve the purpose of identifying different kind of orientational behavior in the system. The distribution profile of $\alpha_m$ could be considered as an order parameter to distinguish any different kind of orientational phase from the complete random disordered state i.e., plastic crystal phase. The role of other ``lower order'' axes and improper operations in defining the unique orientations, will be discussed below.

The highest order rotational symmetry axes of the particles were defined by the set $\mathcal{S}^{p}_{max}$. For the convenience, to represent the axes corresponding to any specific rotational operation, we used another nomenclature. We defined the highest order rotational symmetry axes of the particle point group as $\mathcal{P}_{max}(C_a)$ where $C_a$ was the particular rotational operation with $a$ as the integer number denoting the $a$-\textit{fold} rotation. In principle, the notation $\mathcal{P}_{max}(C_a)$ was the symmetry specific representation for a polyhedral shape of the general description of $\mathcal{S}^{p}_{max}$. In the equivalent way, $\mathcal{P}_{min}(C_a)$ was used to represent ``lower order'' particle axes for the rotational operation $C_a$. We also denoted the rotational axes corresponding to the operations $C_b$ of crystallographic point group as $\mathcal{U}(C_b)$, where the integer number $b$ indicated the $b$-\textit{fold} rotation in the corresponding crystal structure. It followed that $\mathcal{P}_{max}(C_a)$ and $\mathcal{U}(C_b)$ were the elements of the sets $\mathcal{S}^{p}_{max}$ and $\mathcal{S}^{c}$ respectively as defined in section \ref{symm_relation}.

\begin{figure}[!h]
	\centering 
	\includegraphics[scale=0.95]{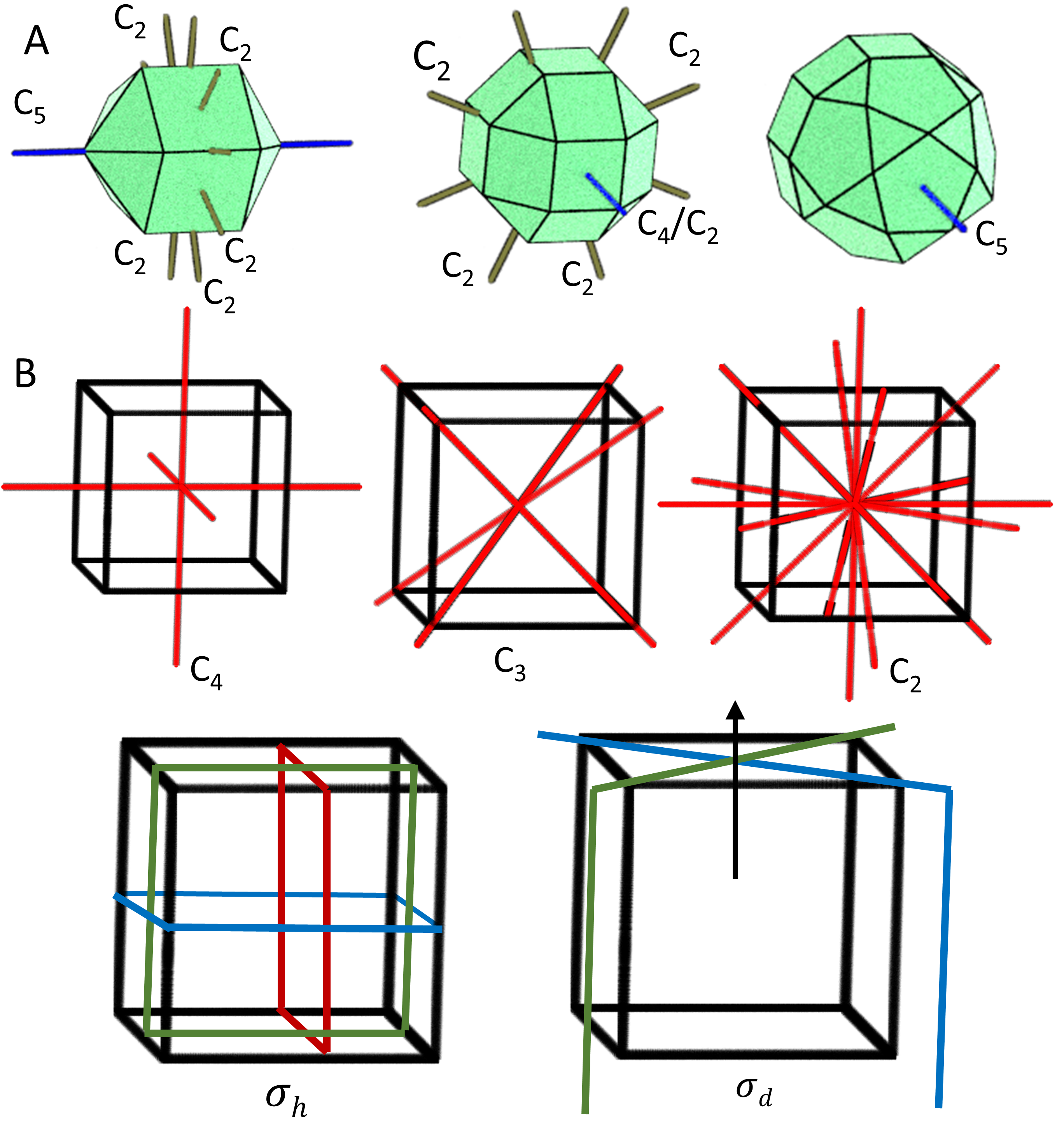}
	\caption{\textbf{Symmetry operations of the particle point groups and crystallographic point group are shown:} \textbf{(A)} the axes corresponding to the proper rotational symmetry operations of three shapes, EPD ($D_{5h}$), ESG ($D_{4d}$) and EPG ($C_{5v}$), \textbf{(B)} the axes of the proper rotational symmetry operations of ``cubic'' crystal along with the $\sigma_h$ and $\sigma_d$ plane in $O_{h}$ point group. The highest order axes corresponding to the particle point groups are shown in ``blue'' color and other ``lower order'' particle axes are shown in ``dark-olive''color.}
	\label{fig:symm_particle_crystal}
\end{figure}

For each shape used in this study, there are only two axes of the particle point group corresponding to the highest order rotational symmetry if two opposite directions of a single axis are considered as non-equivalent explicitly, i.e., the $\mathcal{P}_{max}(C_5)$ axes of EPD (point group - $D_{5h}$), $\mathcal{P}_{max}(C_4)$ axes of ESG (point group - $D_{4d}$) and $\mathcal{P}_{max}(C_5)$ axes of EPG (point group - $C_{5v}$) shapes. Each of the EPD and ESG shapes contains another five axes as ``lower order'' corresponding to the rotational operation in the point groups of $D_{5h}$ and $D_{4d}$ respectively. All the shapes are portrayed in Fig.\,\ref{fig:symm_particle_crystal}A with all axes corresponding to the proper rotational operations in the point groups. It should be noted that all the five ``lower order'' axes of the EPD shape reside in the perpendicular plane of the $\mathcal{P}_{max}(C_5)$ axes. This shape also contains two planes of symmetry i.e., $\sigma_{h}$ and $\sigma_{d}$ in the $D_{5h}$ point group. For the ESG shape, one of the five ``lower order'' axes coincides with the direction of $\mathcal{P}_{max}(C_4)$ axes while the other four axes stay in the perpendicular plane of this axis. The $D_{4d}$ point group contains a $\sigma_{d}$ plane. No ``lower order'' axes corresponding to the rotational symmetry operations exists in the point group of EPG shape ($C_{5v}$). As all three systems self-assemble into FCC crystals, the crystallographic point group appeared to be $O_h$ under the ``cubic'' class consisting of the proper rotational operations 8$C_3$, 6$C_2$, 6$C_4$, 3$C'_2$. The axes of proper rotational operations of the $O_{h}$ symmetry with $\sigma_{h}$ and $\sigma_{d}$ planes are shown in Fig.\,\ref{fig:symm_particle_crystal}B. As these polyhedral shapes crystallize into orientationally disordered FCC crystals, the lattice sites of the unit cells are expected to be occupied by the particles which are orientationally disordered. In general all the lattice sites of the FCC unit cells can be considered as distinguishable barring the possibility of few sites occupied by the particles with same orientations occurring at certain cases. As a result, the number of axes corresponding to the rotational operations $C_3$, $C_2$ $C_4$ are eight, twelve and six respectively. In the $O_h$ symmetry, three axes of $C'_2$ operations coincide with the $\mathcal{P}_{max}(C_4)$ axes.

To observe the discrete hopping among the unique orientations, the idea of transition matrix was used to observe the probability of hopping in the ``discrete rotator phase'' \cite{Kundu2023}. Several previous investigations also reported the existence of this phase in the disordered crystal \cite{Shen2019, Lee2023}. In such phase, particles jump discretely among the finite set of unique orientations exhibited by the disordered crystal. It was expected that the discrete mobility of the particles might involve the highest order rotational axes of the particle point group in such phase. There was also a certain possibility of occurring multiple unique orientations while keeping the direction of an axis constant in three dimensions. Considering all these possibilities, the construction of transition matrix in the context of highest order rotational symmetry axes of the particle, is discussed below. 

\subsection{Transition matrix analysis based on rotational axes} \label{trans_matrix}
After the detection of unique orientations $\mathbb{Q}$, the highest order particle symmetry axes were evaluated using the equation \ref{eq:unique_axes}.
\begin{equation}\label{eq:unique_axes}
	\mathbb{S}^{p}_{ref, k} = \mathbb{Q}_k \mathcal{A}^{p}_{max}
\end{equation}
where $\mathbb{S}^{p}_{ref, k}$ was the unique rotational axes of the particles corresponding to the $k$-\textit{th} unique orientation $\mathbb{Q}_k$. The transition matrix, $\underline{\underline{T}}$, is a $N_{\Omega}\times N_{\Omega}$ matrix characterizing the discrete hopping of particles between the preferred reference axes along a trajectory. The number of transitions or hops between two unique reference axes $\mathbb{S}^{p}_{ref, w}$ and $\mathbb{S}^{p}_{ref, v}$ for the particle $i$, $\mathcal{N}_i(w,v)$ is the number of times the particle toggled between these two reference axes along the trajectory, keeping the order preserved, between two consecutive frames. The case of $w=v$ meant particle did not change axes at all. The elements of particles averaged transition matrix is defined as follows.
\begin{eqnarray}
\underline{\underline{T}}(w,v) = \frac{1}{N}\sum_{i=1}^{N}\frac{\mathcal{N}_i(w,v)}{(l-1)}
\end{eqnarray}
Here, $l$ is the number of frames in the Monte Carlo trajectory. The diagonal elements would give the fraction of jumps onto the same reference axes, i.e.\,an estimate of retention probability, averaged over all particles along the entire trajectory. The population in the off-diagonal elements corresponded the likelihood of hopping between distinct reference axes. If discrete mobility was present, one could expect to see the populations in the off-diagonal elements of the transition matrix. Sufficient frames of the trajectory was chosen to observe the discrete hopping of the highest order rotational axes of the particles and the probability of discrete hopping was quantified, where the discrete rotator phase was observed.

\begin{figure*}
	\centering 
	\includegraphics[scale=0.2]{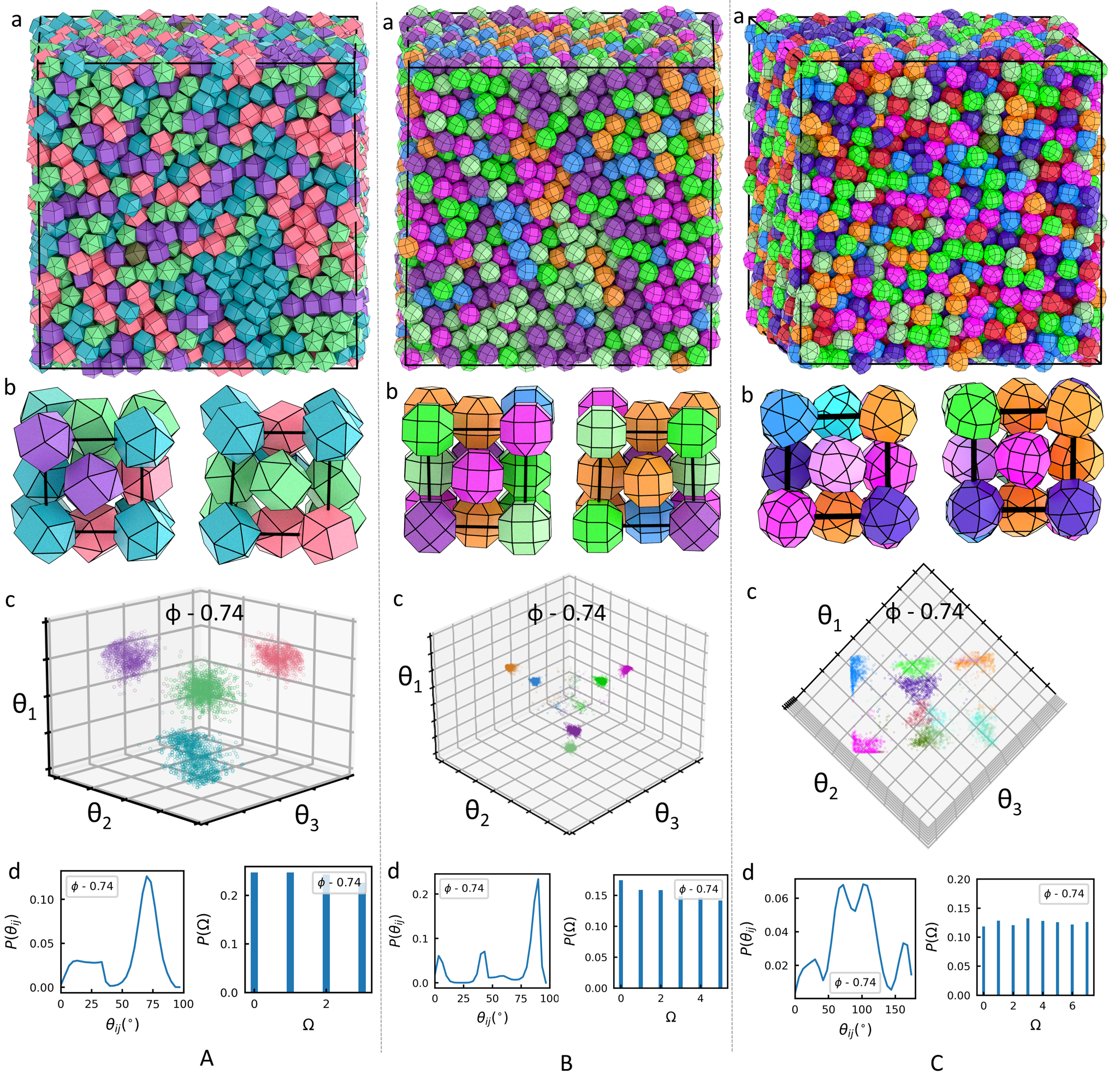}
	\caption{\textbf{Simulation snapshots including crystal unit cells and unique orientations in the systems, distributions of all pairwise angles and population densities of the particles within the unique orientations are shown for three shapes:} \textbf{(A)} EPD, \textbf{(B)} ESG and \textbf{(C)} EPG. Panel (a) of each subfigure depicts the simulation snapshot at the densest packing ($\phi \sim$ 0.74 for all shapes) with the particles shown in different colors based on the unique orientations. Two unit cells of each crystal structure are illustrated in the panel (b) showing particle orientation at each lattice site. The three dimensional space of $\Theta_1$, $\Theta_2$ and $\Theta_3$ exhibiting the unique orientations in the system are shown n panel (c) as the clusters in orientational space. The distributions of all pairwise angles in the system and histograms of population densities within the unique orientations are displayed in the panel (d). These analyses suggest the existence of finite number of unique orientations in each system indicating a few orientational attributes i.e., pairwise orientational differences, equal partitioning of system particles, to be preserved in this phase.}
	\label{fig:snapshots_corr_diosorder}
\end{figure*}

\section{Results}
\subsection{Existence of the multiple unique orientations in the disordered crystals}
The systems of Elongated Pentagonal Dipyramid (EPD), Elongated Square Gyrobicupola (ESG) and Elongated Pentagonal Gyrocupolarotunda (EPG) shapes appeared to exhibit FCC crystals at sufficiently higher pressure regions \cite{Damasceno2012c}. In the crystalline assemblies, all the particles depicted few unique orientations as revealed by our earlier investigation \cite{Kundu2023}. The simulation snapshots of the EPD, ESG and EPG shapes at the highest achievable packing fractions ($\phi$ $\sim$ 0.74 for all shapes) in self-assembly are shown in the panel (a) of Figs.\,\ref{fig:snapshots_corr_diosorder}A,\,B,\,C respectively. Corresponding FCC unit cells of the respective crystal structures are depicted in the panel (b) of Figs.\,\ref{fig:snapshots_corr_diosorder}A,\,B,\,C indicating orientationally disordered crystal following the notion of translationally connected particles to be orientationally disordered \cite{Kundu2023}. In the unit cells and system snapshots, particles were colored based on the unique orientations as showcased in the form of clusters in three dimensional orientational space spanned by $\Theta_1$, $\Theta_2$ and $\Theta_3$ (panel (c) of Figs.\,\ref{fig:snapshots_corr_diosorder}A,\,B,\,C). For each system, a fixed set of colors was used which remained unaltered for the rest of discussion. Each cluster in the three dimensional orientational space which was shown using a fixed color code, represented a particular unique orientation and the particles corresponding to that orientation could be separated out from others. In this setting, total number of clusters indicated the number of unique orientations and orientationally ordered particles within a tolerance angle ($\theta_c$) were classified in each system. The number of unique orientations was four, six and eight for the systems of EPD, ESG and EPG shapes where the values of $\theta_c$ appeared to be 40$^{\circ}$, 20$^{\circ}$ and 45$^{\circ}$ respectively. A detail discussion of this representation was pointed out in our previous investigation \cite{Kundu2023}. The particles were found to preserve fixed orientational differences giving prominent peaks in the distributions of all pairwise angles (first figure of panel (d) of Fig.\,\ref{fig:snapshots_corr_diosorder}). In our previous investigation, we discussed qualitative difference between these distributions and ``shark-fin'' distribution indicating freely rotating plastic crystal phase \cite{Kundu2023}. A detail characterization also revealed equal population densities of particles in the unique orientations as shown in the panel (d) (second figure) of Figs.\,\ref{fig:snapshots_corr_diosorder}A,B,C. 

Though Fig.\,\ref{fig:snapshots_corr_diosorder} depicts the analyses at densest states only, we analyzed the data at other packing fractions lower than the highest ones, in a similar fashion. We restricted ourselves to discuss these data again as this was done previously describing the notion of orientational order/disorder and methodological advances \cite{Kundu2023}.  It was observed that in this phase which was different from freely rotating plastic phase, all systems exhibited frozen orientational states at the densest packing and ``discrete rotator phase'' at comparatively lower packing fractions while the same crystal structure (FCC) was retained in the entire solid regions. In the ``discrete rotator phase'', particles changed their orientations in a discrete way by hopping from one orientation to another such that, number of unique orientations with fixed differences and equipartition of the particles in entire crystalline system remained conserved. Before our investigation, the existence of discrete rotator phase was also reported in the simulations of hard polygons \cite{Shen2019}, colloidal clathrates formed by the hard polyhedra \cite{Lee2023}.

\begin{figure}[!h]
	\centering 
	\includegraphics[scale=1.0]{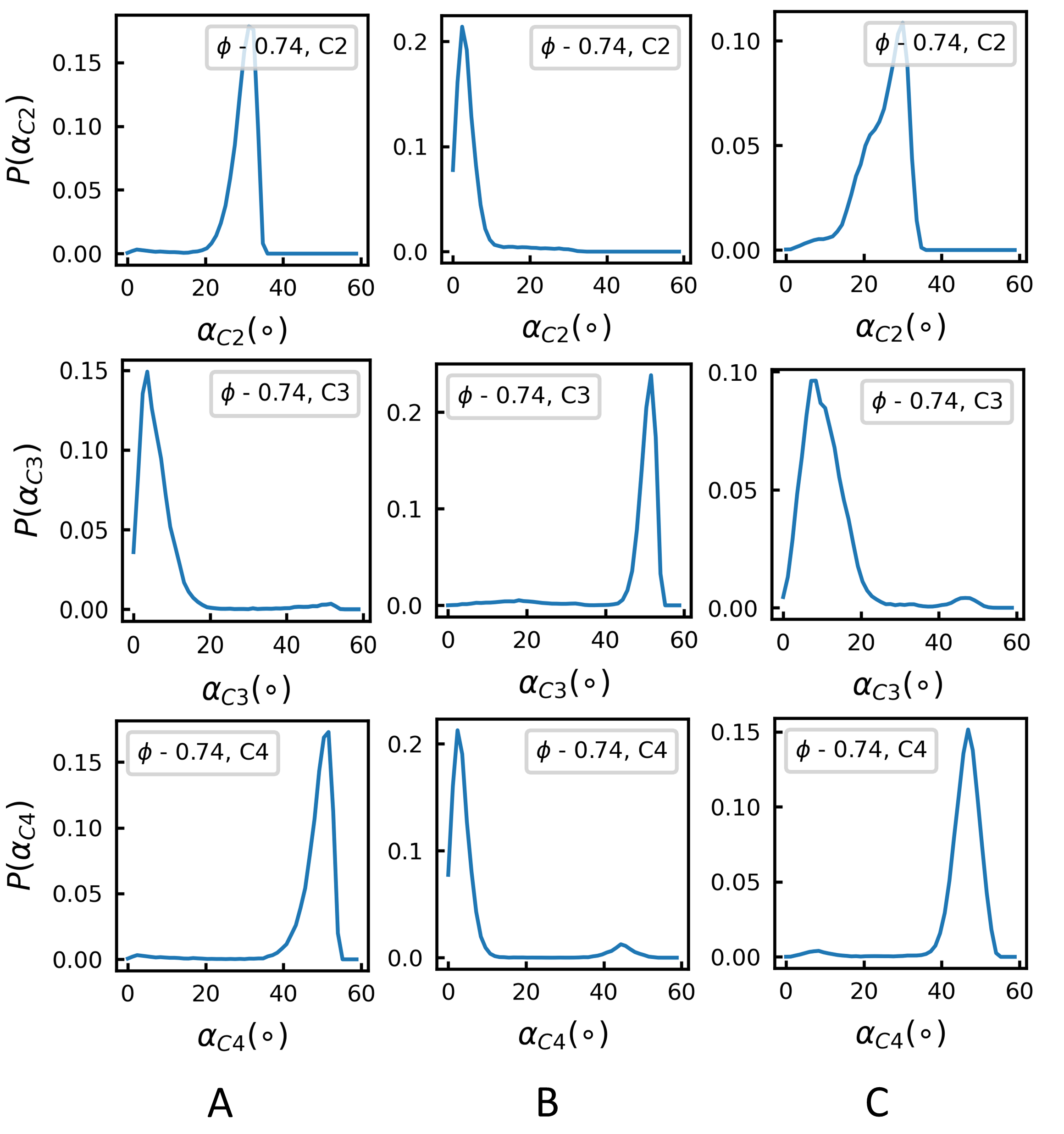}
	\caption{\textbf{Alignment of the $\mathcal{S}^{p}_{max}$ axes and $\mathcal{S}^{c}$ axes for each kind of rotational operation:} \textbf{(A)} EPD, \textbf{(B)} ESG and \textbf{(C)} EPG. For EPD and EPG shapes, the distributions of $\alpha_{C3}$ exhibit the peaks within $20^{\circ}$ indicating the $\mathcal{P}_{max}(C_5)$ axes are parallel with the $\mathcal{U}(C_3)$ axes barring the statistical noise. No parallel alignment are found for $\mathcal{U}(C_2)$ or $\mathcal{U}(C_4)$ axes producing the peaks of the respective distributions away from $0^{\circ}$. For ESG shape, the $\mathcal{P}_{max}(C_4)$ axes are parallel with both the $\mathcal{U}(C_2)$ or $\mathcal{U}(C_4)$ axes of the crystal barring the noise, but the distribution of $\alpha_{C3}$ exhibits a peak at $\sim$ $50^{\circ}$ corresponding to the misalignment of the particles.}
	\label{fig:angle_particle_crystal_per_opration}
\end{figure}

\subsection{Alignment of the rotational symmetry axes of particles and crystals}
To realize the orientational alignments of the particles located at the lattice sites of the FCC unit cells in a more precise way, the particle symmetry and the crystallographic symmetry were taken into account simultaneously. As we investigated earlier, occurrence of multiple unique orientations in the disordered crystal explicitly depended on the unit cells depicted by the crystal structure \cite{Kundu2023}. The highest order axes corresponding to the proper rotational operations of the particle point groups were considered. As the point group of FCC crystal ($O_h$) contains only $C_2$, $C_3$ and $C_4$ rotational operations, the alignments of the highest order rotational symmetry axes of the particles ($\mathcal{S}^{p}_{max}$) were realized through the ensemble averaged distributions of $\alpha_{C2}$, $\alpha_{C3}$ and $\alpha_{C4}$ calculated with $\mathcal{U}(C_2)$, $\mathcal{U}(C_3)$, $\mathcal{U}(C_4)$ crystal axes respectively. For the three shapes EPD, ESG and EPG, the $\mathcal{S}^{p}_{max}$ axes turned out to be $\mathcal{P}_{max}(C_5)$, $\mathcal{P}_{max}(C_4)$, $\mathcal{P}_{max}(C_5)$ respectively while represented by the specific rotational operations as the notations were introduced in the section \ref{repn}. In the distributions of $\alpha_{C2}$, $\alpha_{C3}$ and $\alpha_{C4}$, the sets of axes $\mathcal{U}(C_2)$, $\mathcal{U}(C_3)$, $\mathcal{U}(C_4)$, were considered as indistinguishable within the same set. So, the axes corresponding to the $C_2$ and $C'_2$ operations were considered as equivalent. Both the EPD and EPG shapes appear to contain the $\mathcal{P}_{max}(C_5)$ axes. For these two shapes, the distributions of $\alpha_{C3}$ exhibited the peaks around $\sim$ $10^{\circ}$ indicating the alignments to be parallel at the densest states ($\phi$ $\sim$ 0.74 for both cases) barring the statistical noise as shown in the second row of Figs.\,\ref{fig:angle_particle_crystal_per_opration}A,\,C. The other two distributions (of $\alpha_{C2}$  and $\alpha_{C4}$) displayed the peaks at $\sim$ $30^{\circ}$ and $\sim$ $50^{\circ}$ respectively by ruling out the possibility of parallel alignments of the $\mathcal{P}_{max}(C_5)$ axes of the particles with $\mathcal{U}(C_2)$ and $\mathcal{U}(C_4)$ axes. Similar kinds of analysis were implemented for ESG shapes at $\phi$ $\sim$ 0.74 and the data are shown in Fig.\,\ref{fig:angle_particle_crystal_per_opration}B. This suggests the alignments of the $\mathcal{P}_{max}(C_4)$ axes were parallel with both the $\mathcal{U}(C_2)$ and $\mathcal{U}(C_4)$ axes exhibiting the peaks at $\sim$ $5^{\circ}$ for both cases. For this shape, the distribution of $\alpha_{C3}$ exhibited a peak at $\sim$ $50^{\circ}$ suggesting the misalignment of the $\mathcal{P}_{max}(C_4)$ axes with the $\mathcal{U}(C_3)$ axes. We also investigated the effect of all ``lower order'' axes of these polyhedral shapes to find any direct correspondence with the $\mathcal{U}(C_2)$, $\mathcal{U}(C_3)$, $\mathcal{U}(C_4)$ axes in the self-assembled crystal structures. The distributions of angles $\alpha_{C2}$, $\alpha_{C3}$ and $\alpha_{C4}$ did not indicate any parallel alignments for the $\mathcal{P}_{min}(C_2)$ axes of both EPD and ESG shapes which explicitly ruled out the possibility of any other axes except the ``highest order'' in the alignments of the particles (see SI Fig.\,11 for details). As the EPG shape does not contain any ``lower order'' axes, this analyses were not carried out for this shape.

In the presence of multiple unique orientations, these analyses provided the information of parallel arrangements between the $\mathcal{S}^{p}_{max}$ axes and $\mathcal{S}^{c}$ axes for any specific rotational operation $C_s$. The $\mathcal{P}_{max}(C_5)$ axes of EPD and EPG shapes were observed to be in parallel with the $\mathcal{U}(C_3)$ axes, where as, the $\mathcal{P}_{max}(C_4)$ axes of ESG shapes were parallel with the $\mathcal{U}(C_2)$ and $\mathcal{U}(C_4)$ axes of the crystal. It should be noted, we found four and eight unique orientations for the EPD and EPG shapes respectively as shown in Figs.\,\ref{fig:snapshots_corr_diosorder}A,C. For these two shapes the parallel alignments occurred with eight $\mathcal{U}(C_3)$ axes. Similarly, for the ESG shape, the number of unique orientations was six as shown in Fig.\,\ref{fig:snapshots_corr_diosorder}B. But both the $\mathcal{U}(C_4)$ and $\mathcal{U}(C_2)$ axes were responsible for ESG shapes in the orientational alignments of the particles. The existence of finite set of unique orientations of these bodies will be justified later in the context of parallel alignments between the $\mathcal{S}^{p}_{max}$ and $\mathcal{S}^{c}$ axes.

\begin{figure*}
	\centering 
	\includegraphics[scale=0.95]{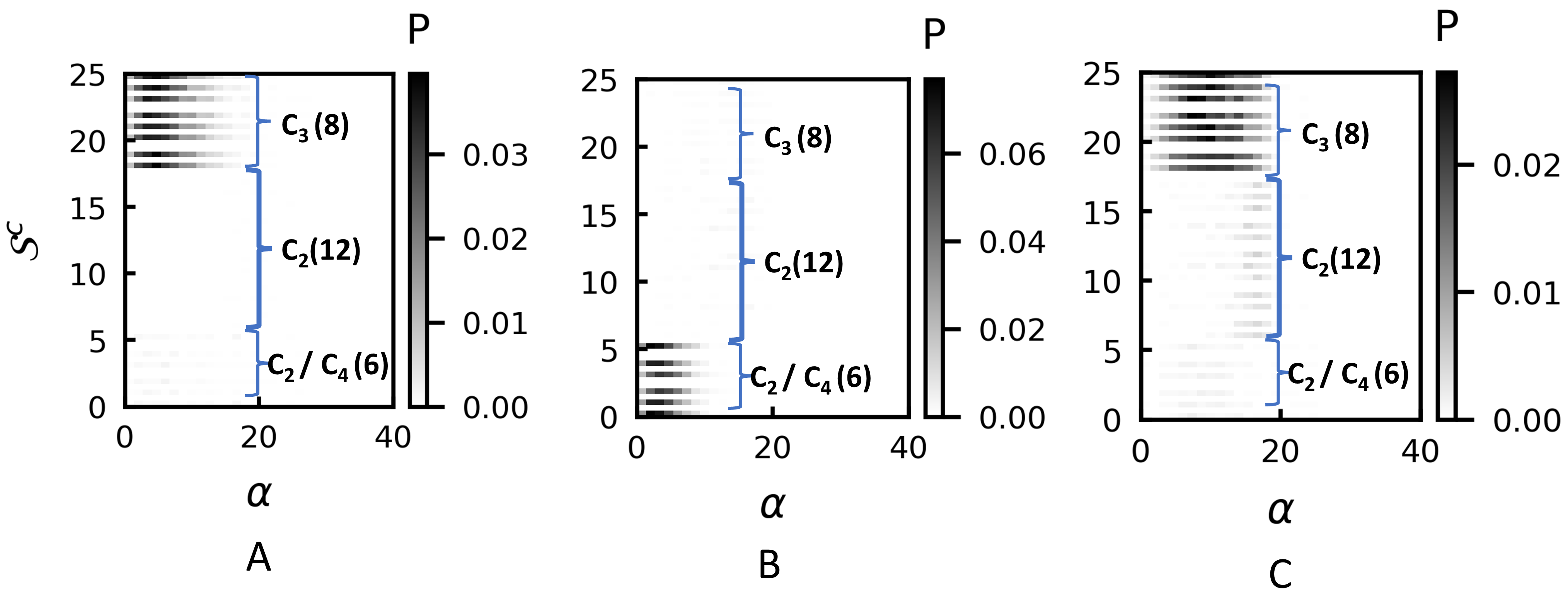}
	\caption{\textbf{The distributions of $\alpha$ calculated between the axes of $\mathcal{S}^{p}_{max}$ and $\mathcal{S}^{c}$ for three shapes at highest packing fractions:} \textbf{(A)} EPD and \textbf{(B)} ESG and \textbf{(C)} EPG. For EPD and EPG shapes, the $\mathcal{P}_{max}(C_5)$ axes form $\sim$ $10^{\circ}$ with eight  $\mathcal{U}(C_3)$ axes of the $O_h$ crystallographic point group indicating the parallel alignments occur with equal probabilities. For ESG shape, the $\mathcal{P}_{max}(C_4)$ axes are found to be parallel in the directions of six $\mathcal{U}(C_4)$ axes of the crystal. All the particles in the system become equally partitioned within the finite set of axes corresponding to particular rotational operation of the crystallographic point group.}
	\label{fig:angle_particle_crystal_all}
\end{figure*}

Our next analyses included the study of two-dimensional histograms of $\alpha$ for three shapes. In this distribution, all $\mathcal{S}^{c}$ axes were considered as distinguishable and indexed by integers. The ensemble averaged histograms of $\alpha$ calculated between the axes of the sets $\hat{\mathcal{S}}^{p}_{max}$ and $\hat{\mathcal{S}}^{c}$ were investigated at the densest states ($\phi$ $\sim$ 0.74 for all cases). In principle, the distribution of $\alpha$ appeared as an alternate representation of the plots shown in Fig.\,\ref{fig:angle_particle_crystal_per_opration}. But a major difference between these plots lied in identifying the distinguishable crystallographic axes based on the operation $C_s$. So, by construction, these histograms were expected to provide more microscopic details than the previous analyses. In the distributions shown in Fig.\,\ref{fig:angle_particle_crystal_all}A,\,B,\,C, all the crystallographic symmetry axes ($\mathcal{S}^{c}$) were categorized in such way, so that $\mathcal{S}^{c}$ $\in$ (0, 5), $\mathcal{S}^{c}$ $\in$ (6, 17) and $\mathcal{S}^{c}$ $\in$ (18, 25) corresponded to $\mathcal{U}(C_4)$, $\mathcal{U}(C_2)$ and $\mathcal{U}(C_3)$ axes of $O_h$ point group respectively, where the three $C'_2$ axes were not shown separately as those coincided with the $C_4$ axes. For EPD and EPG shapes, the $\mathcal{U}(C_3)$ axes of the crystal were responsible in the parallel alignments with $\mathcal{P}_{max}(C_5)$ axes forming the angles within $10^{\circ}$ as indicated in Figs.\,\ref{fig:angle_particle_crystal_all}A,\,C. The color bar represents the normalized population density at each crystal axis. It was noticeable that all the eight $\mathcal{U}(C_3)$ axes were equally populated by the particles in the system barring the statistical noise which also showed the negligible effect of any other axes in the crystal i.e., $\mathcal{U}(C_2)$ and $\mathcal{U}(C_4)$. As Fig.\,\ref{fig:angle_particle_crystal_all}B suggests, the $\mathcal{P}_{max}(C_4)$ axes of ESG shapes appeared to be aligned with the six $\mathcal{U}(C_4)$ axes where all the particles were also equally compartmentalized within the six directions satisfying the appearance of six unique orientations in the ESG system. This two-dimensional histogram analysis revealed the exact reason of the appearance of a peak  at $\sim$ $0^{\circ}$ in the distribution of $\alpha_{C2}$ as shown in Fig.\,\ref{fig:angle_particle_crystal_per_opration}B. It showed, as the three axes corresponding to the $C'_2$ operations in $O_h$ point group were not considered separately, a peak apparently came indicating the parallel alignments of the ESG bodies with the $\mathcal{U}(C_2)$ axes also. From the histogram shown in Fig.\,\ref{fig:angle_particle_crystal_all}B, it was convincible that for ESG shapes the $\mathcal{P}_{max}(C_4)$ axes made parallel alignment with six $\mathcal{U}(C_4)$ axes of the FCC crystal. As the analysis suggested, the alignments of the rotational axes between the highest order particle symmetry and crystallographic symmetry turned out to be a characteristic feature of the phase where multiple unique orientations occurred in a disordered crystalline structure. In the same way, the distributions of $\alpha$ in the plastic crystal phases were observed for the three shapes which exhibited almost Gaussian distribution showing the distinct differences in the orientational arrangements of particles in the plastic crystal and this phase, as shown in SI Fig.\,7.

\begin{figure*}
	\centering 
	\includegraphics[scale=0.2]{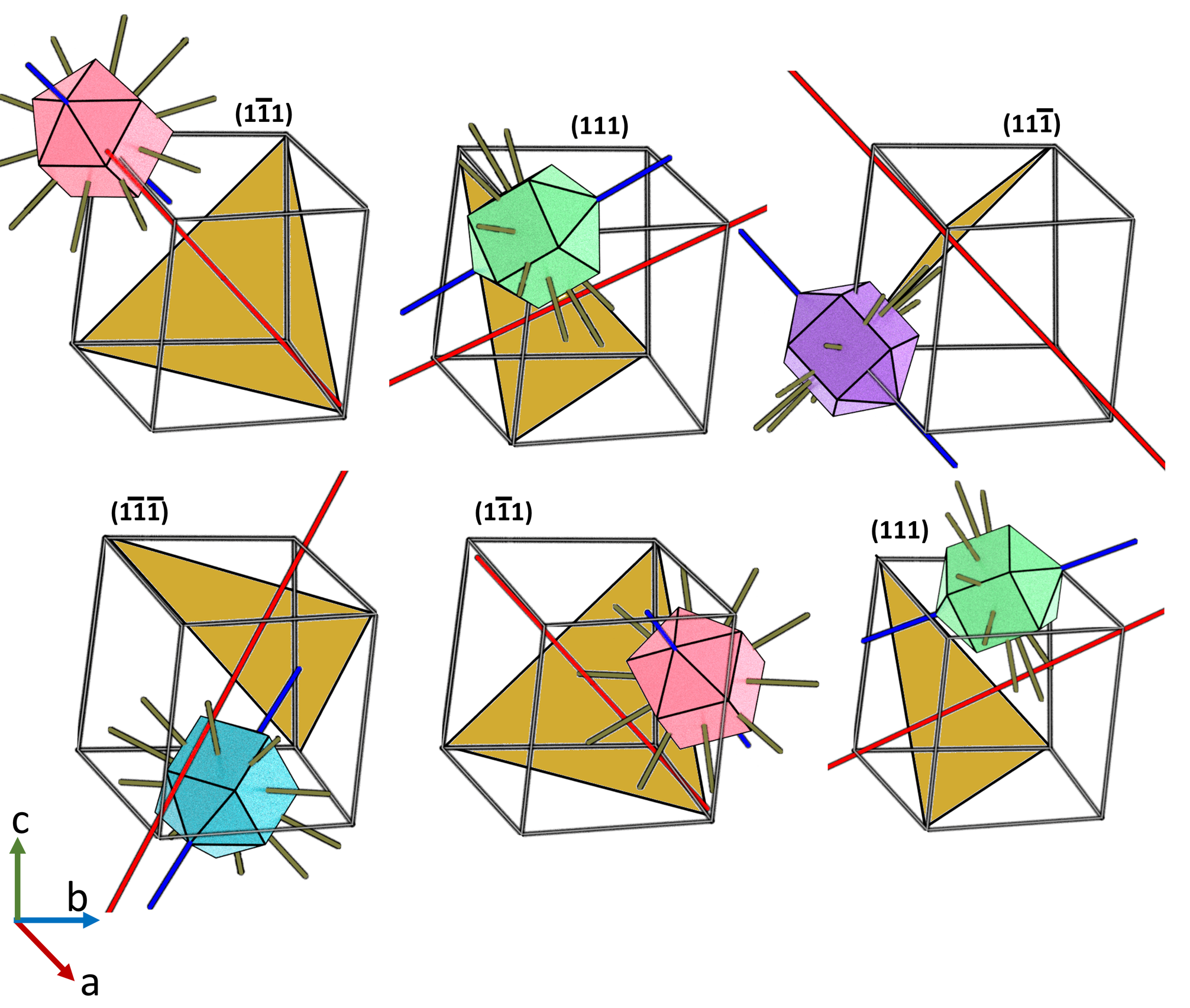}
	\caption{\textbf{The alignments of the particles in the FCC unit cell are shown for EPD shapes.} A randomly chosen FCC unit cell with three corner particles and three face particles are portrayed. The highest order particle symmetry axes are shown in ``blue'' color and other lower order axes are represented as ``dark-olive'' color. The corresponding crystallographic rotational symmetry axes responsible for the parallel alignments are represented by ``red'' color. The $\mathcal{P}_{max}(C_5)$ axes of the particles are found to be parallel with the $\mathcal{U}(C_3)$ axes of the crystal. The (\textsl{hkl}) planes are also shown in ``yellow'' color (also marked in text) for the unit cell which are perpendicular to the $\mathcal{P}_{max}(C_5)$ directions. The  lattice translational vectors are shown in the bottom-left corner of the figure.}
	\label{fig:EPD_axes}
\end{figure*}

\begin{figure*}
	\centering 
	\includegraphics[scale=0.23]{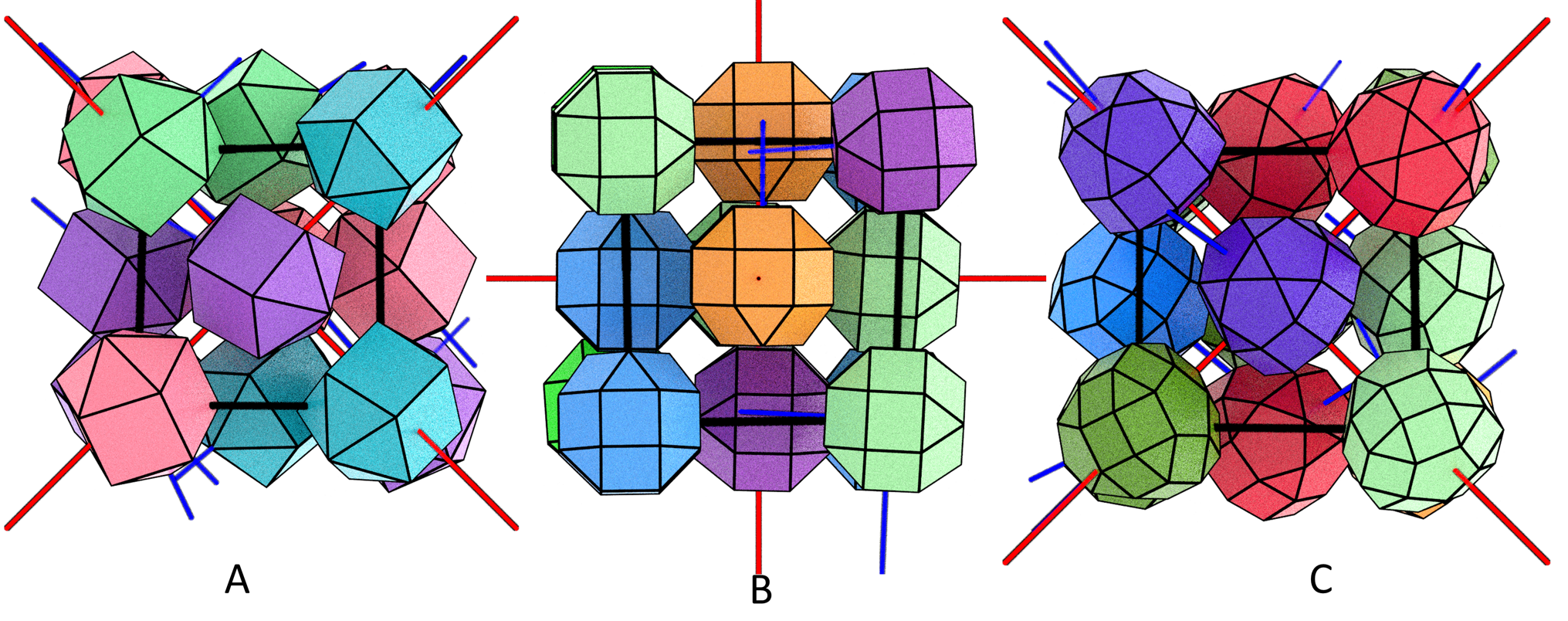}
	\caption{\textbf{A random unit cell from the FCC crystalline structure at $\phi$ $\sim$ 0.74 are shown for each of the three shapes:} \textbf{(A)} EPD and \textbf{(B)} ESG and \textbf{(C)} EPG. The particle's highest order rotational symmetry axes in the ``blue'' color and corresponding crystallographic rotational axes in the ``red'' color. For EPD and EPG shapes, the $\mathcal{P}_{max}(C_5)$ axes are parallel with any of the $\mathcal{U}(C_3)$ axes shown in the red color. For ESG system, $\mathcal{P}_{max}(C_4)$ are parallel with any of the six $\mathcal{U}(C_4)$ axes of the crystal. The color codes of the particles are kept fixed for all systems as defined in Fig.\,\ref{fig:snapshots_corr_diosorder}.}
	\label{fig:summary}
\end{figure*}

The visual representation of the alignment of $\mathcal{P}_{max}(C_5)$ axes of EPD shapes in a randomly chosen unit cell of FCC crystals is shown in Fig.\,\ref{fig:EPD_axes} at $\phi$ $\sim$ 0.74. Fig.\,\ref{fig:EPD_axes} displays total six particles out of fourteen in the FCC unit cell; three corner and three face particles. It was important to notice the orientations of all fourteen particles in the unit cell and not the four effective particles only. All the EPD particles located at fourteen lattice sites of the same unit cell are shown in SI Fig.\,8 separately. It also gave the essence of orientationally disordered crystal as argued in our previous study \cite{Kundu2023}. The EPD shape contains $D_{5h}$ point group and the $\mathcal{P}_{max}(C_5)$ axes of the body made parallel alignment with the $\mathcal{U}(C_3)$ axes of the crystal as illustrated by the Fig.\,\ref{fig:EPD_axes}. The color codes of the polyhedral particles were kept fixed and all rotational symmetry axes of the body are shown along with shape itself at the proper lattice site. At each subfigure, only two $\mathcal{U}(C_3)$ axes (separated by $180^{\circ}$) of the crystal are displayed in the unit cell which were responsible to make the parallel alignment; $\alpha$ $\sim$ $0^{\circ}$ or $\sim$ $180^{\circ}$. The corresponding (\textsl{hkl}) planes of the crystal are also displayed and marked as text. A randomly chosen FCC unit cell for each of the three shapes are shown in Fig.\,\ref{fig:summary}. It suggests a common feature of this phase which appear in the form of the alignment of highest order particle symmetry axes with any rotational symmetry axes of the crystal structure. The data of the other two FCC forming shapes ESG and EPG, are presented in SI Figs.\,9,10 in the similar fashion. Our observation regarding the straightforward relationship between the particle and crystal symmetry in determining the discrete set of unique orientations in the disordered crystals appear to be a common source of the inter-particle orientational correlation which preserves the systemwise unique orientations with fixed angular difference and equal partitioning of all particles.

To explore the orientationally ordered phase in the context of the symmetry relationship between the particle and crystal, we performed the simulations of Truncated Cuboctahedron (TC) shape which was reported to form orientationally ordered body-centred tetragonal (BCT) structure with two unique orientations at the densest state \cite{Karas2019, Kundu2023}. It was known that the self-assembled BCT crystal of TC shape contains the particles two unique orientations with the orientational difference of $\sim$ 45$^{\circ}$. As TC shape contains $O_h$ point group symmetry, six $C_4$ axes appears to be the highest order rotational axes (considering two axes separated by 180$^{\circ}$ as non-equivalent). In the orientationally ordered structure, no such rotational symmetry axes of crystallographic point group was found to be responsible in the parallel alignment of the particles defining the orientational order (see SI Fig.\,12 for details). In the orientationally ordered crystal structure, multiple unique orientations with fixed orientational differences existed with equal population densities, but no such symmetry relationship between the particle and crystal appeared s a characteristics phenomenon indicating the essential difference of this phase with orientationally ordered crystal.

\begin{figure*}
	\centering 
	\includegraphics[scale=0.95]{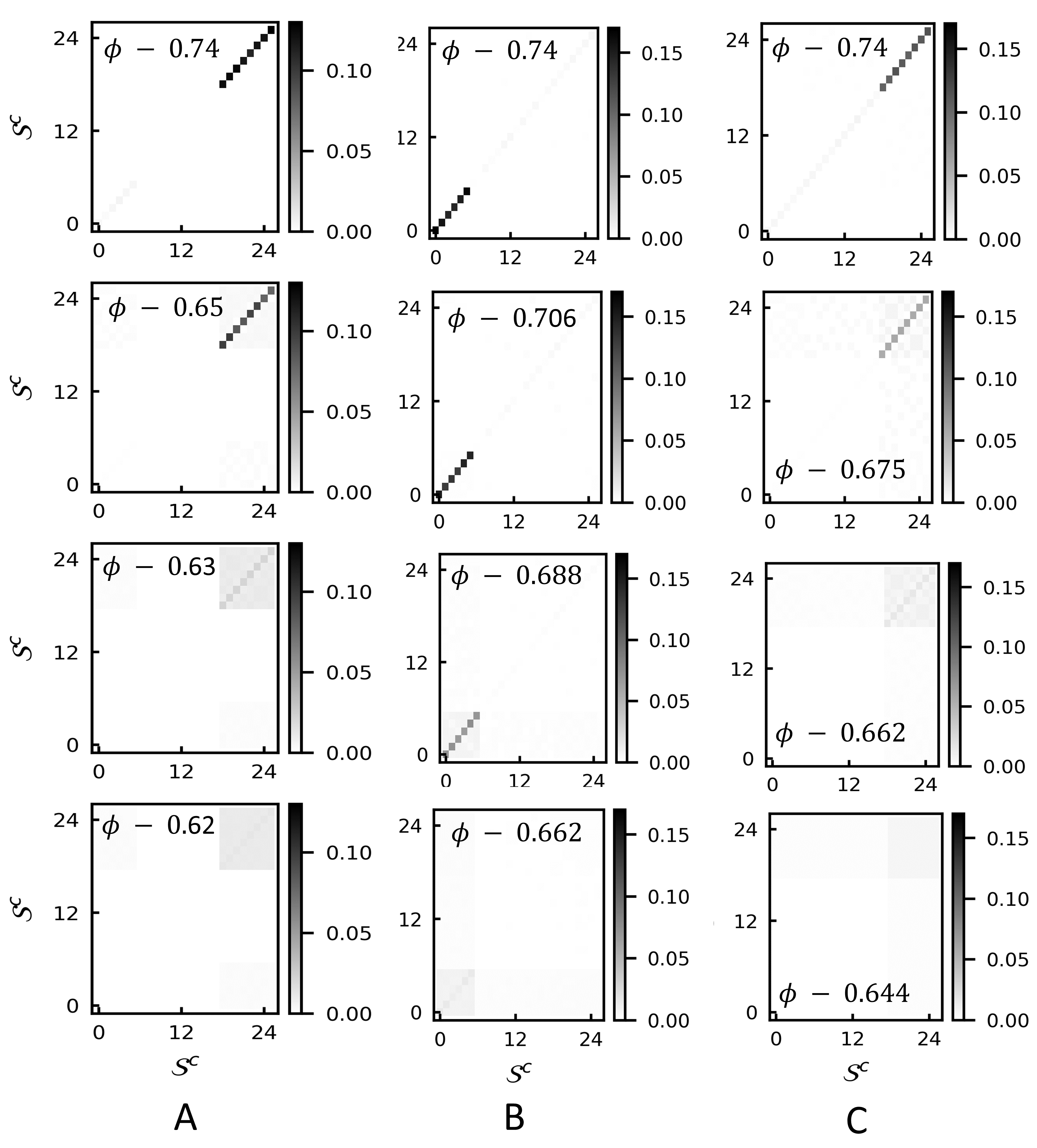}
	\caption{\textbf{Discrete directional hopping of the highest order symmetry axes of the particles within the fixed set of crystallographic rotational axes are shown for the three shapes:} \textbf{(A)} EPD, \textbf{(B)} ESG and \textbf{(C)} EPG. The top row suggest the frozen motion of the particles leading to the population at only the diagonal elements of the transition matrix. The off-diagonal elements gets populated as the discrete changes of the particles' rotational axes are observed upon melting of the systems. The last row suggests the homogeneous distributions at lower packing fractions where the directional hopping occurs with equal probability for all particles.}
	\label{fig:hopping}
\end{figure*}

\subsection{Discrete mobility of the highest order rotational axes of the particles}
In the discrete rotator phase, three shapes; EPD, ESG and EPG, exhibited orientational hopping within the unique orientations depending on the room allowed for the particles when the systems were melted slightly from the densest states \cite{Kundu2023}. At the highest achievable packing fractions, the orientational motion of all particles were frozen within a finite set of discrete orientations while maintaining the translational order depicted by the crystal structures. The similar signatures were already reported in the systems of hard polygons \cite{Shen2019} and colloidal clathrates \cite{Lee2023}. This discrete orientational behavior persisted before the arrival of the plastic crystal phase with complete freely rotating particles. One possible interpretation of the orientational hopping could be anticipated as the discrete directional change of highest order rotational axis of the particles. As the orientational hopping always occurred within the unique orientations, it was expected that the highest order particle symmetry axes would toggle within a few fixed directions. To investigate this phenomena, we implemented the analyses of transition matrix to capture the probabilities of hopping within fixed set of directions. In this analyses also, the crystallographic symmetry axes ($\mathcal{S}^{c}$) were classified in such way, so that $\mathcal{S}^{c}$ $\in$ (0, 5), $\mathcal{S}^{c}$ $\in$ (6, 17) and $\mathcal{S}^{c}$ $\in$ (18, 25) corresponded to $\mathcal{U}(C_4)$, $\mathcal{U}(C_2)$ and $\mathcal{U}(C_3)$ axes of $O_h$ point group respectively. The three $C'_2$ axes of of $O_h$ point group were not shown separately as those coincided with the $C_4$ axes. Each value in the two dimensional matrix indicated the hopping probability of $\hat{\mathcal{S}}^{p}_{max}$ axes within the respective crystallographic axes $\mathcal{S}^{c}$ which were responsible for the alignments of the corresponding particles. The orientational motion of the EPD and EPG shapes were frozen at the highest packing fractions ($\phi$ $\sim$ 0.74) resulting in a state with no rotational mobility of the $\hat{\mathcal{S}}^{p}_{max}$ axes. This gave rise to the equal population densities at the eight diagonal elements of the transition matrices barring the statistical noise, as shown in the first row of Figs.\,\ref{fig:hopping}A, C. The previous analyses suggested, the $\mathcal{P}_{max}(C_5)$ axes of EPD and EPG bodies aligned with the $\mathcal{U}(C_3)$ axes of the crystal (see Figs.\,\ref{fig:angle_particle_crystal_all}A,C). Confirming the previous results, the data of transition matrix indicated the probability of discrete directional changes of $\mathcal{P}_{max}(C_5)$ axes gradually increased as the packing fraction decreased from the densest state. At $\phi$ $\sim$ 0.65 and 0.675 for EPD and EPG shapes respectively, maximum particles were found without the discrete mobility, while very less population were observed in the off-diagonal elements indicating the $\mathcal{P}_{max}(C_5)$ axes of very few particles exhibited discrete hopping. A distinct differences between the numbers of frozen particles and hopping particles was observed at these packing fractions. As the rotational mobility of the particles increased, more number of particles started to hop discretely within the fixed set of $\mathcal{U}(C_3)$ axes resulting in a very less difference in the numbers of hopping particles and non-hopping particles. This difference almost tended to zero at $\phi$ $\sim$ 0.62 and 0.644 for the EPD and EPG shapes respectively, where the $\mathcal{P}_{max}(C_5)$ axes of all particles changed the directions with equal probabilities ignoring the statistical noise. As a result, homogeneous distributions were noticed in the transition matrices within the specific region dedicated to the $\mathcal{U}(C_3)$ axes with no discernible differences in the population densities between the diagonal and off-diagonal elements. Similarly, for ESG shape, the $\mathcal{P}_{max}(C_4)$ axes chose six finite directions of $\mathcal{U}(C_4)$ axes and no discrete mobility was observed at $\phi$ $\sim$ 0.74 as shown in Fig.\,\ref{fig:hopping}B. At $\phi$ $\sim$ 0.688, a majority of population remained frozen at the corresponding orientational states showing the population densities only at the diagonal elements. If the packing fraction decreased further, the $\mathcal{P}_{max}(C_4)$ axes of the particles started to hop discretely within the six specific axes. The number of frozen particles became almost zero at $\phi$ $\sim$ 0.662, where all particles exhibited discrete hopping within six $\mathcal{U}(C_4)$ axes with equal probabilities resulting in a homogeneous distribution in the transition matrix barring the statistical noise. For the ESG shape, we failed to observe the two-step correlation in the transition matrices as discussed in the earlier research. The absence of the two-step correlation will be discussed later in support of our observations. This analyses suggest that, even in the discrete rotator phase, where the highest order rotational symmetry axes of the particle discretely toggled within a fixed set of crystallographic axes with the equal probabilities, the number of unique orientations and equal compartmentalization of the particles remain constant in the entire crystal structure. 

\section{Discussion}
We report the computational evidence of a direct relationship between the particle point group and crystallographic point group in controlling the multiple unique orientations in the disordered crystals in the discrete rotator phase. Though the symmetry of local environment was associated with the polyhedral particle to rationalize the discrete orientations in earlier studies \cite{Shen2019, Lee2023}, our observation suggest a more robust affinity explaining the unique orientations of the particles incorporating the symmetry of the crystal structure. As the orientational disorder was maintained in the entire system without any spatial arrangements, the straightforward expectation would suggest the occurrence of unique orientations following certain relationships with the unit cell of the crystal. We considered all previously reported observations and define a common feature exhibited by the disordered structures in the presence of multiple unique orientations. This symmetry relationship confirmed the existence of long-range orientational correlation in the entire crystal structure due to the occurrence at the level of unit cells. This non-trivial affinity between the symmetries of particle and crystal structure indicated the conservation of a ``local rule'' as found in the disordered ice crystal according to Bernal-Fowler rule \cite{Bernal1933}. In entropy driven assembly, the discrete rotator phase appeared be completely symmetry protected where this direct connection between the point groups of two entities could be the source of correlation and other orientational attributes such as the number of unique orientations with fixed differences, equal population density per unique orientation, discrete orientational hopping were the signatures of the correlation as discussed in detail in the following paragraphs. The existence of orientational correlation in the disordered crystalline solids could lead to the possibility of interpreting the ``discrete rotator phase'' as ``correlated orientational disorder'' as observed in the earlier reports of multi-component atomic crystals \cite{Keen2015, Simonov2020, Meekel2021}.

In this research, we explained the existence of multiple unique orientations and discrete hopping of the particles from the symmetry aspect of both the particle and crystal structure. In the EPD system, the number of unique orientations appeared to be four (see second row of Fig.\,\ref{fig:snapshots_corr_diosorder}A) where as the $\mathcal{P}_{max}(C_5)$ axes of the shape aligned in the directions of eight $\mathcal{U}(C_3)$ axes. To explain four unique orientations of EPD shape in terms of the symmetry relationship, proper rotational operations and symmetry of the mirror planes present in the particle point group are required to be considered. As the point group of EPD shape ($D_{5h}$), contains a $\sigma_{h}$ plane, the shape looks identical about this plane which is perpendicular to the $\mathcal{P}_{max}(C_5)$ axes. In this case, two $\mathcal{P}_{max}(C_5)$ axes separated by $180^{\circ}$ can be considered as equivalent. As a result, the number of unique orientations turns down to four from eight for EPD shape. This argument is straightforward and can be verified for the EPG shape containing $C_{5v}$ point group with no $\sigma_{h}$ plane leading to the existence of eight unique orientations in the system. This is consistent with the symmetry relationship as $\mathcal{P}_{max}(C_5)$ axes of EPG shape aligned in the parallel directions with eight $\mathcal{U}(C_3)$ axes of the crystal structure. For the ESG shape with $D_{4d}$ point group, the number of unique orientations appeared to be six and the $\mathcal{P}_{max}(C_4)$ axes of the particle were parallel with the six $\mathcal{U}(C_4)$ axes of the $O_h$ symmetry. For this shape, two distinct unique orientations can not be considered as equivalent considering the symmetry of the mirror plane present in the respective point group. We show the evidences that all the unique orientations occurred in such way, it broke certain symmetries of the corresponding crystal structure which could be realized by combining the point groups of both the particle and crystal. Moreover, the fixed angular differences among the particular kind of rotational axes of the crystallographic point group suggest the fixed difference among the unique orientations. Nevertheless, one possible interpretation of the equal compartmentalization of the particles within the unique orientations arises from the equivalence of all the axes corresponding to any particular kind of rotational operation in the crystallographic point group. If the alignments between the particle and crystallographic rotational symmetry axes are preserved at the unit cells of the crystals, then the existence of (i) finite number of unique orientations, (ii) fixed orientational differences, (iii) equipartition of the particles in the system can be interpreted as the direct signatures of such relation. The existence of long-range behavior can also be explained in terms of the occurrence of this direct affinity at the level of unit cells leading to the manifestation of this phenomena in the entire crystalline system. This relationship appeared to be independent of the system size which could not be possible to explain with any statistical arguments. This indicates an obvious orientational correlation in the form of symmetry relationship between the particle and crystal, only two entities existed in the system under the influence of hard-core interaction. At slightly lower packing fractions, the ``discrete rotator phase'' could be the outcome of this correlation exhibiting discrete hopping within the unique orientations following the conservation of all signatures. The data of TC shape proved the fact that only the existence of multiple orientations would not lead to the possibility of this phase satisfying the symmetry relationship. As our analyses suggests the crystalline system would correspond to orientationally disordered structure in the presence of multiple discrete orientations to exhibit the correlation. We speculate that this symmetry protected phenomena could be the only reason for the occurrence of the phase for some polyhedral shapes which crystallized into the structures with different point groups from the corresponding shapes.

In the ESG system, there was a clear signature of second order correlation while investigating the discrete orientational hopping at lower packing fractions as reported in the previous study \cite{Kundu2023}. We found no such signatures in the transition matrix for the ESG shapes upon melting of the system as shown in Fig.\,\ref{fig:hopping}B. The  possible argument suggests that the presence of unique orientations separated by $45^{\circ}$ (second row of  Fig.\,\ref{fig:snapshots_corr_diosorder}B) does not affect the direction of $\mathcal{P}_{max}(C_4)$ axes of $D_{4d}$ point group. Alternatively, it can be justified in this way, two distinct orientations of the ESG shapes separated by the quaternion angle $\theta$ $\sim$ $45^{\circ}$ leads to the ``axis-angle'' difference of one $\mathcal{P}_{max}(C_4)$ axis by $\alpha$ $\sim$ $180^{\circ}$. The definition of quaternion angle $\theta$ and ``axis-angle'' $\alpha$ were introduced in the sections \ref{unique_orein} and \ref{symm_relation} respectively. Following this argument, if the hopping between two unique orientations separated by $45^{\circ}$ occurs, no change in the direction of $\mathcal{P}_{max}(C_4)$ axes will be observed in the transition matrix. Consequently, no signature of second order correlation will be observed as displayed in Fig.\,\ref{fig:hopping}B. The existence of second order correlation depends on the polyhedral shape itself and does not appear to be a generic feature of the orientational correlation. We justify that consideration of both translational order and rotational symmetry elements of the crystallographic point group along with all symmetry elements of particle point group, may be enough to describe all the exposed signatures as uncovered in the study. 

In the orientationally ordered phase of TC shape, no straightforward relationship could be drawn combining the rotational symmetry elements of the particle point group with crystallographic point group. From our investigations, the following conclusions can be inherited: (1) orientational ordered phase consists of highest order rotational symmetry axes $\mathcal{S}^{p}_{max}$ of the  particle creating any finite angles with the rotational symmetry axes $\mathcal{S}^{c}$ of the crystal, (2) in the discrete rotator phase, the $\mathcal{S}^{p}_{max}$ axes of the particles make parallel alignments with the $\mathcal{S}^{c}$ axes of the underlying crystal structure and (3) in the plastic crystal phase, the particles rotate freely following no such constraints imposed by the underlying crystal structure. However, we did not observe the effect of other attributes except the particle and crystal symmetry responsible for the conservation of various orientational signatures in the entropy driven assembly. The current investigation authorizes the combined role of both kind of symmetries up to a certain extent, which turns out to be a general characteristics exhibited by all particles in the system and acts like the source of long-range orientational correlation. Our observation reveals that single particle symmetry can be coupled with the crystalline symmetry to decipher the complex nature of the orientational correlation which can lead to the fundamental understanding of orientational disorder in the crystalline assemblies.

Though, we reported the data in the crystal structures of three shapes with multiple unique orientations which appeared as a generic feature of this phase, our investigation may trigger a few obvious questions. In this study, it turns out to be any particular kind of axes of the respective crystallographic point group (any of the $C_4$, $C_3$ and $C_2$ operation of $O_h$ symmetry) were responsible in the alignment of the particles. But we are not able to justify any specific reasons behind the involvement of a particular kind of crystallographic axes. In general, the alignments may occur with multiple kinds of such axes if large number of unique orientations exist in the system and the possibilities are open for further studies. Again, all the shapes have only two highest order rotational symmetry axes in the respective point group where two axes are separated by $180^{\circ}$. But, the effect of symmetry relationship is not reported for those polyhedra with multiple choices of such axes which are not exactly parallel. Moreover, we did not observe these phenomena in any other crystal structures except $O_h$ symmetry. All these questions will be addressed in further investigations, which can facilitate to enhance the fundamental understanding of orientational disorder in crystalline assemblies.

\section{Conclusion}
Our analyses reveal a different aspect to study the unique orientations in the disordered crystalline structure which includes the analyses of single particle point group combining with the point group symmetry of the corresponding crystal structure. We establish the role of translational and rotational symmetry of the crystal structure in controlling such disordered phase with long-range orientational correlation in the entropy driven assembly. A similar kind of signatures were already mentioned in the molecular crystals where the phase was termed as ``orientational glass'' phase \cite{Nitta1959, Loidl1989}. The phase was reported when particle symmetry did not match with the crystal symmetry and the essence of this phase was required to understand from the dynamical point of view. In the entropy driven assembly, we observed the existence of identical signatures in the hard particle assembly and the phase appeared to be in equilibrium phase from every aspect. The existence of ``discrete rotator phase'' also indicates the equilibrium nature ruling out the possibility of any orientational glassy behavior. The freely rotating particles in the plastic phase started to obey some symmetry relationship with the crystal structure when the system transitioned into such phase where multiple but few specific unique orientations existed. Our previous analyses showed the evidences of first-order phase transition between the two phases in these system \cite{Kundu2023}. Our investigations can facilitate to enhance the fundamental understanding of the disordered phase in the simplest model system which may be useful as a design principle of the targeted assembly or to realize the detail implications of correlation in the presence of translational broken symmetry states.

\begin{acknowledgments}
We wish to acknowledge financial support from DST-INSPIRE Fellowship provided to SK. AD thanks DST-SERB Ramanujan Fellowship (SB/S2/RJN-129/2016) and IACS start-up grant. KC thanks IACS for financial support. Computational resources were provided by IACS HPC cluster and partial use of equipment procured under SERB CORE Grant No.\,CRG/2019/006418.
\end{acknowledgments}

\bibliography{crys_symm}

\newpage
\onecolumngrid
\section{Supporting Information}
\subsection{Distribution of $\alpha$ in the plastic phase}
\begin{figure}[!h]
	\centering 
	\includegraphics[scale=0.85]{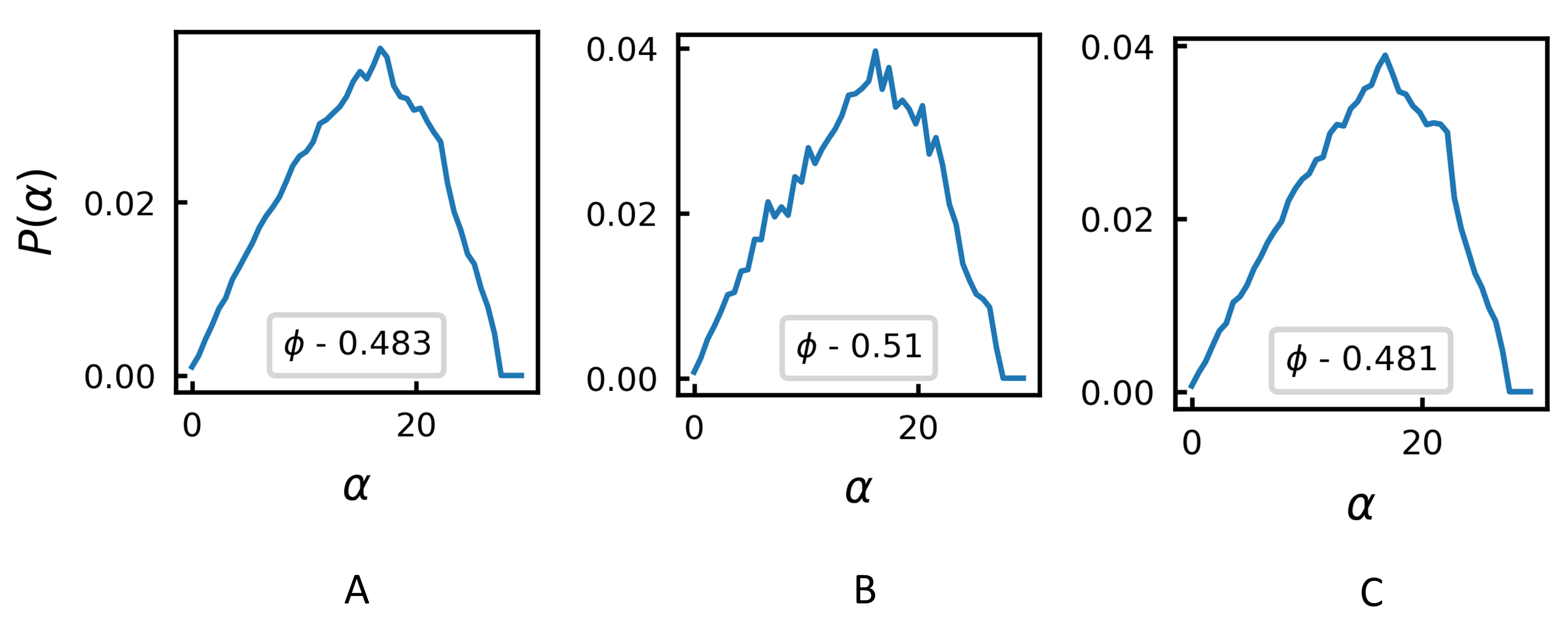}
	\caption{\textbf{The distributions of $\alpha$ in the plastic phase are shown for three shapes:} \textbf{(A)} EPD, \textbf{(B)} ESG and \textbf{(C)} EPG. All the distributions show Gaussian nature as the highest order particle symmetry axes are not restricted to follow any relationship with the rotational symmetry axes of the crystal structures.}
	\label{fig:plastic_phase}
\end{figure}

The distributions of $\alpha$ calculated between the highest order rotational symmetry axes of the particles and all rotational symmetry axes of the crystal are shown in Figs.\,\ref{fig:plastic_phase}A,B,C for the three shapes, EPD, ESG and EPG. All the systems exhibit FCC crystals in the plastic phase where no orientational correlation is observed leading to the Gaussian distributions barring the statistical noise.

\newpage
\subsection{Orientation of the particles and relation with crystal symmetry}
\begin{figure*}
	\centering 
	\includegraphics[scale=0.23]{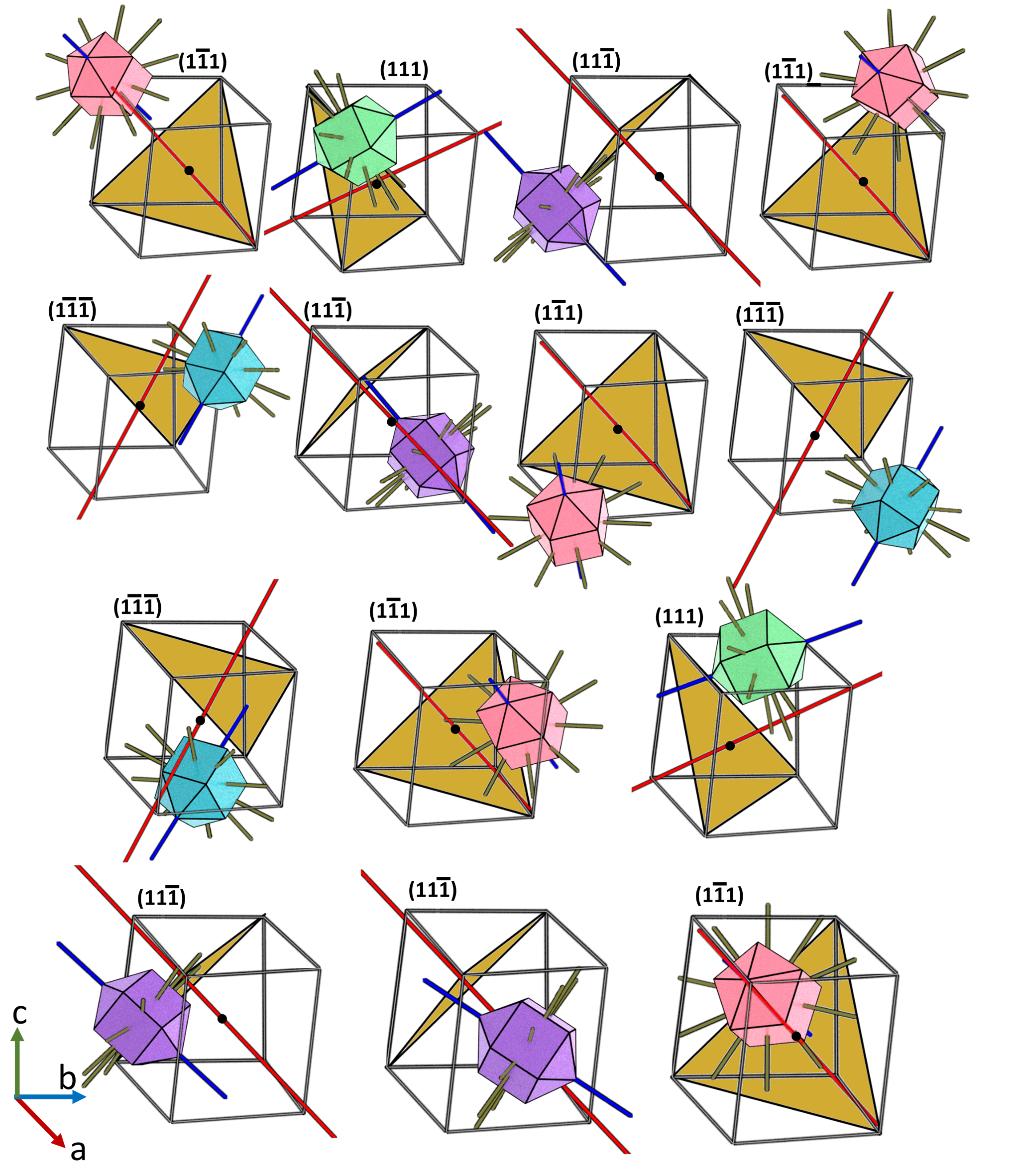}
	\caption{A FCC unit cell consisting of fourteen particles is shown with a single EPD particle at the lattice site of the unit cell in each of the fourteen sub-figures. Only $C_3$ axes of the crystal which are aligned with the $C_5$ symmetry axes of the particle are shown. The (\textsl{hkl}) planes are also shown in the unit cells perpendicular to the $C_3$ axes of the FCC crystal.}
	\label{fig:EPD}
\end{figure*}

\begin{figure*}
	\centering 
	\includegraphics[scale=0.23]{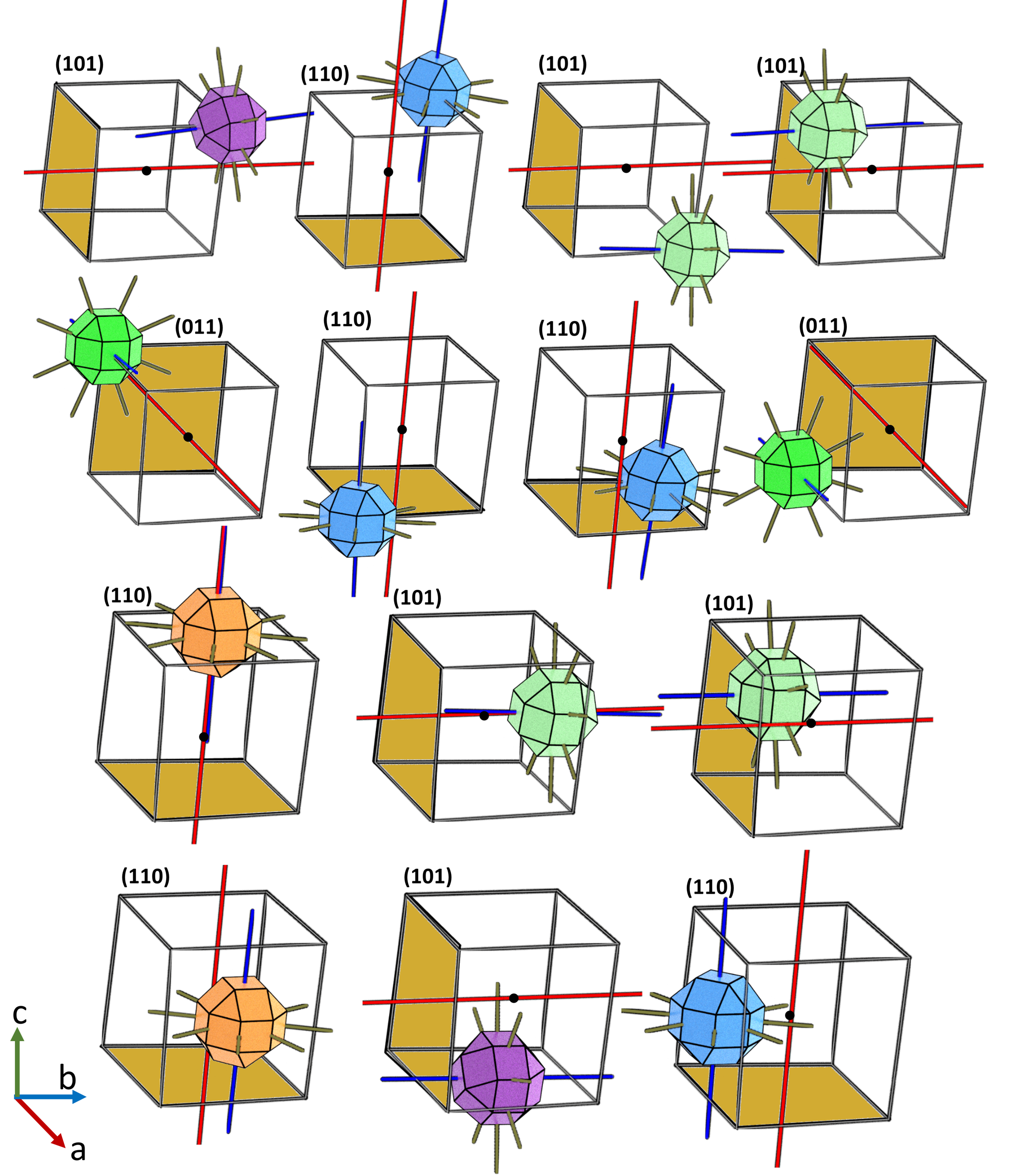}
	\caption{A FCC unit cell consisting of fourteen particles is shown with a single ESG particle at the lattice site of the unit cell in each of the fourteen sub-figures. Only $C_4$ axes of the crystal which are aligned with the $C_4$ axes of the particle are shown. The (\textsl{hkl}) planes are also shown in the unit cells perpendicular to the $C_4$ axes of the FCC crystal.}
	\label{fig:ESG}
\end{figure*}

\begin{figure*}
	\centering 
	\includegraphics[scale=0.23]{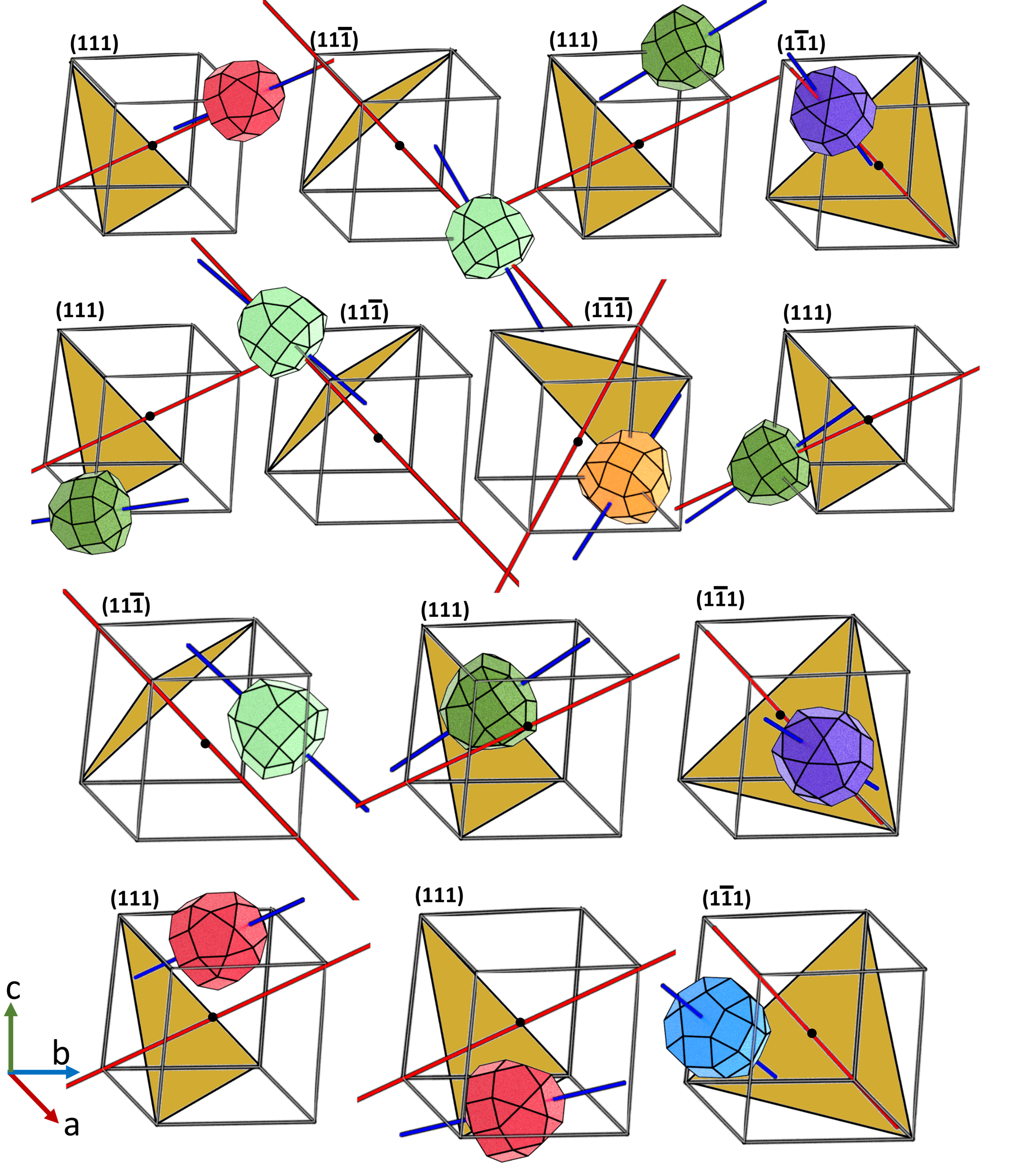}
	\caption{A FCC unit cell consisting of fourteen particles is shown with a single EPG particle at the lattice site of the unit cell in each of the fourteen sub-figures. Only $C_3$ axes of the crystal which are aligned with the axes of $C_5$ symmetry of the particle are shown. The (\textsl{hkl}) planes are also shown in the unit cells perpendicular to the $C_3$ axes of the FCC crystal.}
	\label{fig:EPG}
\end{figure*}

The fourteen particles of a random chosen FCC unit cell from the system of EPD, ESG and EPG are shown separately in fourteen sub-figures of Figs.\,\ref{fig:EPD}, \ref{fig:ESG}, \ref{fig:EPG} respectively; each sub-figure contains a single particle at the proper lattice site with similar color code for better visualization. Fig.\,\ref{fig:EPD} shows that the $C_5$ symmetry axes (the highest order rotational axis of $D_{5h}$ symmetry) of each of the fourteen EPD particles are aligned in the directions of any eight $C_3$ axes of FCC crystal barring the statistical noise. The highest order symmetry axes ($C_4$) of ESG particles with $D_{4d}$ point group also align with six $C_4$ axes of the FCC unit cells as per Fig.\,\ref{fig:ESG}. Similarly, The $C_5$ symmetry axes of each EPG particle staying at the lattice site coincide with the direction of any eight $C_3$ axes of the crystal as shown in Fig.\,\ref{fig:EPG}. The (\textsl{hkl}) planes are also shown in the unit cells which are responsible for the alignment of the particles. The color codes of all particles and axes were followed as defined in the main text.

\newpage
\subsection{Alignments of the lower order rotational symmetry axes of the particles with the axes of crystals}
\begin{figure*}[!h]
	\centering 
	\includegraphics[scale=0.9]{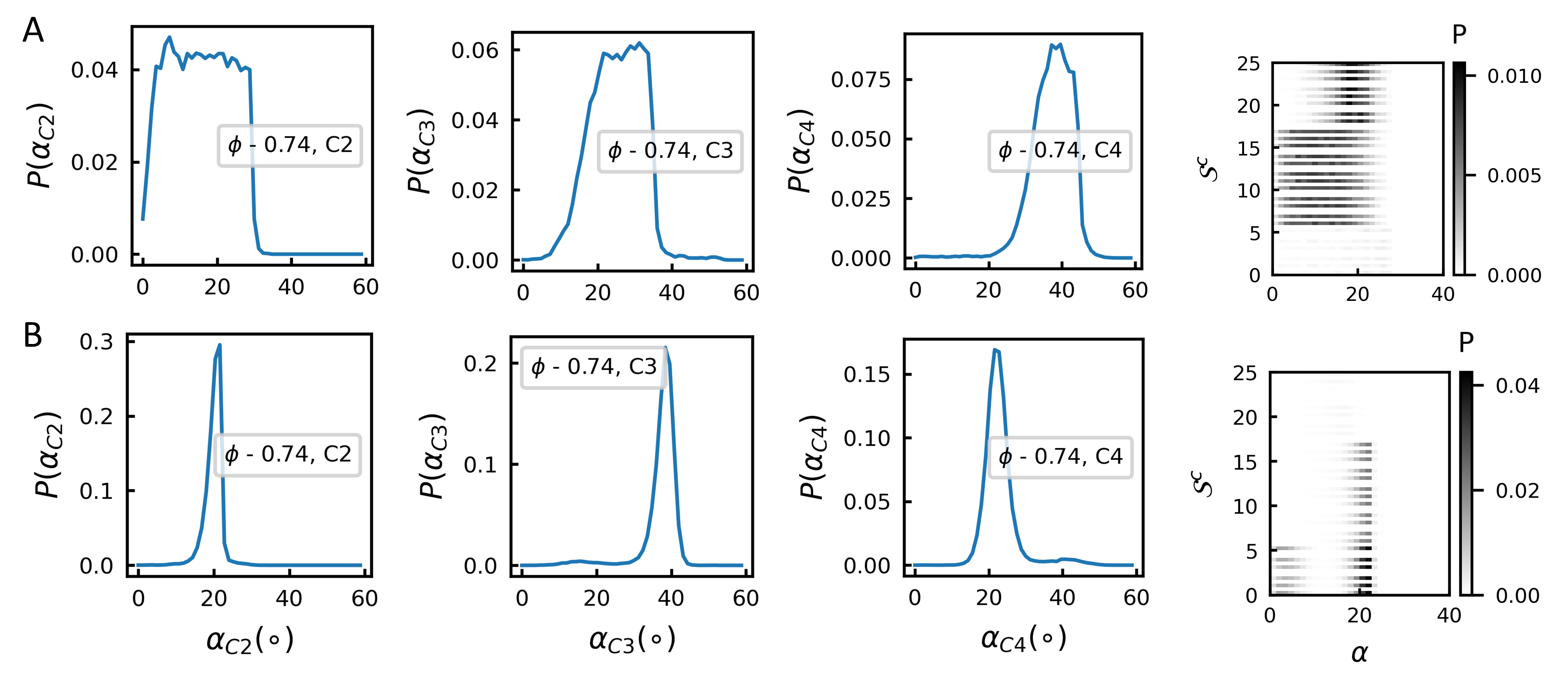}
	\caption{\textbf{The alignments of the lower order rotational symmetry axes of the particle point group are shown at the highest packing fractions for two shapes:} \textbf{(A)} EPD and \textbf{(B)} ESG.  The distributions of angles $\alpha_{C2}$, $\alpha_{C3}$ and $\alpha_{C4}$, estimated between the $C_2$ axes of the particles and $C_2$, $C_3$ and $C_4$ axes of the crystal indicate no direct relationship among the lower order rotational symmetry axes of the particle and all rotational axes of the crystal. The two dimensional distributions of $\alpha$ shown at the extreme right of each sub-figure, does not indicate any linkage of the lower order axes of the particle with crystal structure.}
	\label{fig:lower_order}
\end{figure*}

We explored the effect of lower order rotational symmetry axes of the particles in the alignment of the particles in the unit cells to get the insight about the orientational correlation present in the system. The EPD and ESG shapes contain the $C_5$ and $C_4$ axes respectively as the highest order rotational symmetry axes, but both the shapes have multiple $C_2$ axes present in the respective point groups as the lower order symmetry. The EPG shape contains only the axes of $C_5$ symmetry without having any other rotational symmetry operation in the point group. Figs.\,\ref{fig:lower_order}A,B shows the distributions of $\alpha_{C2}$, $\alpha_{C3}$ and $\alpha_{C4}$ calculated between any of the $C_2$ axis of the particle and $C_2$, $C_3$ and $C_4$ axes of the crystal separately for both shapes. The distributions suggest no direct relationship of the lower order rotational symmetry axes of the particle with the crystal structures where multiple crystal symmetry axes are involved without aligning in the directions of the particle axes.

\newpage
\subsection{Orientational ordered phase of Truncated Cuboctahedron (TC) system}
\begin{figure*}
	\centering 
	\includegraphics[scale=0.25]{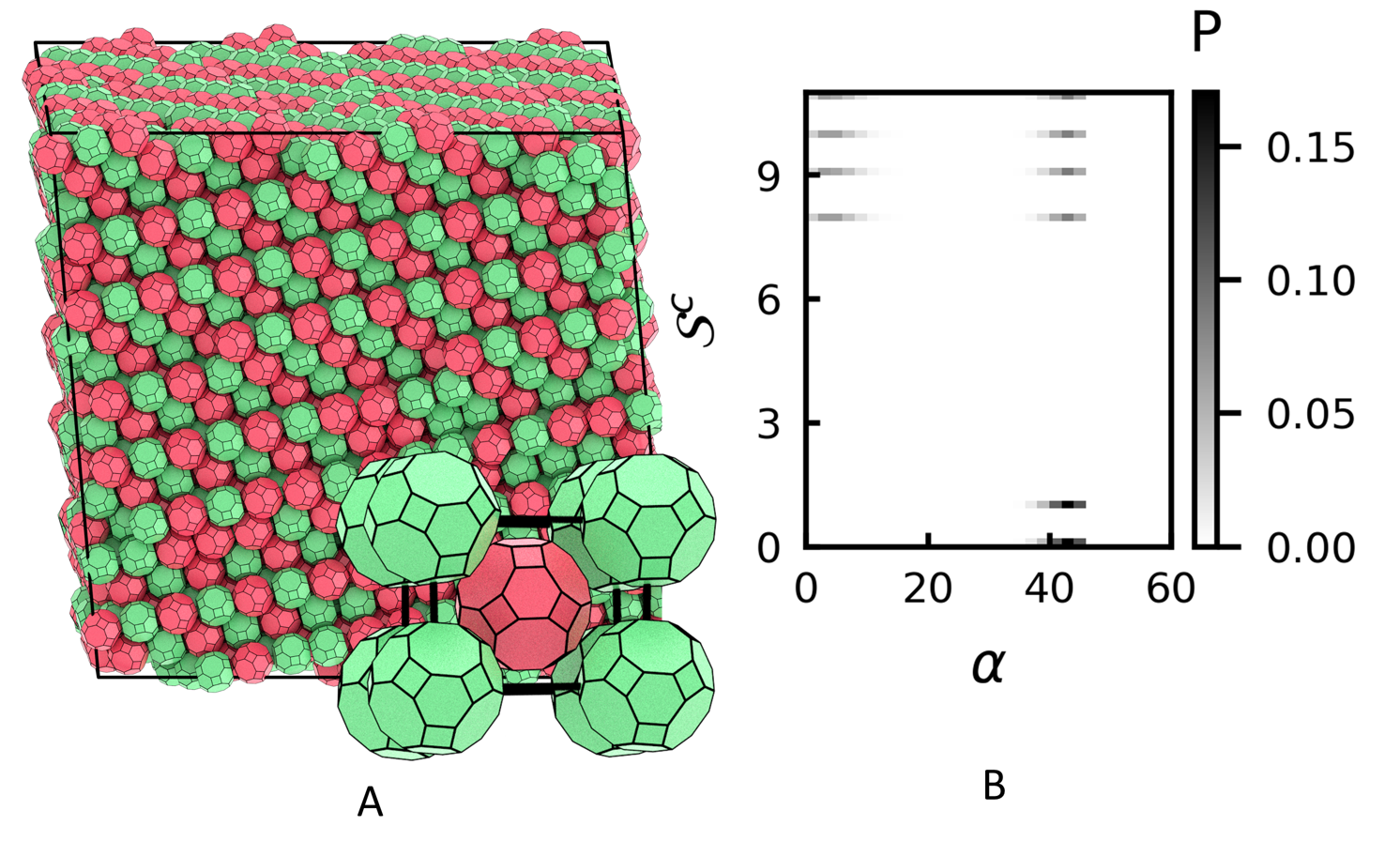}
	\caption{\textbf{The self-assembled BCT crystalline solid of TC shape and the alignment of the particles in the unit cell are shown}. (A) The system snapshot with two unique orientations of TC shapes is shown in different colors in the BCT crystal. A BCT unit cell is shown as inset. The two dimensional histogram of $\alpha$ indicates the appearance of two angles at $\sim$ $10^{\circ}$ and $45^{\circ}$ for all the particles in system as shown in (B). It suggests the non-alignment of the particles in the unit cell while analyzing in the light of the point groups of both particles and the crystal structure.}
	\label{fig:TC}
\end{figure*}

Truncated Cuboctahedron (TC) shape was reported to self-assembly in body centred tetragonal (BCT) structure with tetragonal symmetry of crystallographic point group. In the densest state ($\phi$ $\sim$ 0.74), the self-assembled TC system had two unique orientations with a fixed pairwise quaternion angle difference as shown in Fig.\,\ref{fig:TC}A. The TC shape contains six $C_4$ axes as the highest order rotational axes of $O_h$ point group and the tetragonal symmetry of the crystallographic point group consists of few rotational operations i.e., 2$C_4$, $C_2$, 2$C_2^{'}$, 2$C_2^{''}$. The two dimensional distribution of $\alpha$ estimated between the six $C_4$ axes of the particle and any crystallographic rotational symmetry axes, shows the population density appears at $\sim$ $45^{\circ}$ with the $C_4$ axes of the tetragonal crystal. Other particles in the system are found to create angles $\alpha$ $\sim$ $10^{\circ}$ and $45^{\circ}$ with the $C_2$ crystal axes as shown in Fig.\,\ref{fig:TC}B, where all the $C_2$, $C_2^{'}$, $C_2^{''}$ axes are considered as equivalent. The existence of two unique orientations in the system leaves the signatures of population densities in the histogram at two different $\alpha$ values as confirmed by the Fig.\,\ref{fig:TC}B. This indicates that all the highest order particle symmetry axes are not required to align in the directions of any crystallographic rotational axes to maintain the orientationally ordered crystalline phase as confirmed by the appearance of the peaks at $\alpha$ $\sim$ $45^{\circ}$ shown in Fig.\,\ref{fig:TC}B.

\end{document}